\documentclass[12pt, draftcls, onecolumn]{IEEEtran}
\usepackage{url,amsmath,graphicx}

   \usepackage{cite}
   \usepackage{url}
\usepackage{fancyhdr}
\usepackage{cite}
\usepackage{graphicx}
\usepackage{psfrag}
\usepackage{subfigure}
\usepackage{url}
\usepackage{stfloats}
\usepackage{amsmath}
\usepackage{array}
\usepackage{fancyhdr}
\usepackage{epsfig}
\usepackage{amssymb}
\usepackage{color}
\begin{document}
\title{Unified Stochastic Geometry Modeling and Analysis of  Cellular Networks in LOS/NLOS and Shadowed Fading }
\author{Im\`ene~Trigui, Member, \textit{IEEE}, Sofi\`ene~Affes, Senior Member, \textit{IEEE}, and~Ben Liang, Senior Member, \textit{IEEE}
 }

 \maketitle

  \begin{abstract}
 Statistical characterization of the signal-to-interference-plus-noise ratio (SINR) via its cumulative distribution function (CDF) is ubiquitous in a vast majority of technical contributions
in the area of cellular networks since it boils down to averaging the Laplace transform of the aggregate interference, a benefit accorded at the expense of confinement to the simplistic Rayleigh fading.  In this work, to capture diverse fading channels that arise in
realistic outdoor/indoor wireless communication scenarios, we tackle the problem differently.
By exploiting the moment generating function (MGF) of the SINR, we succeed in analytically assessing cellular networks performance over
the  shadowed $\kappa$-$\mu$,  $\kappa$-$\mu$, and  $\eta$-$\mu$ fading models. The latter
offer high flexibility by capturing diverse fading channels including  Rayleigh, Nakagami-$m$, Rician, and
Rician shadow fading distributions. These channel models have been recently praised for their capability to accurately model dense urban environments, future femtocells, and device-to-device (D2D) shadowed channels.
In addition to unifying the analysis for different channel models, this work integrates  the coverage,  the achievable rate,  and the bit error probability (BEP) which are largely treated separately in the literature. The developed model and analysis are validated over a broad range of simulation  setups and parameters.

 \end{abstract}
\section{INTRODUCTION}
Cellular networks modeling and analysis is a  vibrant topic that keeps taking new dimensions in complexity as to mirror the evolution if not revolution of wireless networks
from the first to the upcoming fifth wireless technology generation (5G). As a key enabler
to realize 5G wireless
networks, heterogeneous networks (HetNets) are indeed the most influential solution that guarantees higher data
rates and  macrocell traffic  off-loading, while providing dedicated capacity to homes, enterprises, or urban
hot spots.
 To cope with such evolution, stochastic geometry proved to be a very powerful tool for reproducing
large-scale spatial randomness, an intrinsic property of emerging cellular networks, as well as different sources of
uncertainties (such as multipath fading, shadowing, and power control) within tractable and accurate mathematical frameworks \cite{andrews},\! \cite{sawy1}.
Able to provide  insightful design guidelines, through closed forms,   stochastic geometry  rid system-level performance evaluation
of computationally-intensive simulations.

In the last decade, many contributions spearheaded this line of research by developing all aspects
of the stochastic geometry models, except for the fading environments.
 For instance, the downlink
baseline operation of cellular networks is characterized in 
\cite{andrews}-\!\!\cite{renzo2}.  Range
expansion and load balancing are studied in \cite{liang},\!\cite{halim}.  By exploiting recent advances in stochastic geometry analysis, several
mathematical frameworks are developed to study multiple-input multiple output (MIMO) operation in cellular networks \cite{MIMO1},\!\cite{MIMO2}. Other aspects including energy
efficiency, energy harvesting,  interference cancellation,  additional interference imposed via underlay
device-to-device (D2D) communication, etc.,  have been investigated  exploiting the tractability of stochastic geometry ( cf.\!\cite{sawy} and references therein).

As far as the fading model is concerned, the Rayleigh fading has been commonly assumed, with only some proposals incorporating the
Nakagami-$m$ fading,  yet merely with integer parameter values \cite{renzo2},\!\cite{leila}.
Such particular fading distributions, by leading to exponential expressions
for the conditional SINR  that enable averaging via the MGF of the interference, have very often implied very similar mathematical
models  in their analysis steps. Strikingly, due to Rayleigh assumption,
 characterizing the SINR via its cumulative distribution function (CDF) is ubiquitous in almost all pioneering contributions pertaining to cellular networks modeling \cite{andrews}-\!\!\cite{sawy}.

Such infatuation with Rayleigh and Nakagami-$m$ has, however, limited  legitimacy according to \cite{kmu},\!\cite{beaulieu}, who argued that these fading
models may fail to capture  new and more realistic fading environments. Besides  ignoring  the line-of-sight (LOS) component in
the received signal, which is prominent in outdoor cellular
communications, the Rayleigh model is a single-parameter fading model
that is not flexible enough to accurately represent complex indoor fading
environments. The diagnosis for Rayleigh fading is even more pessimistic in future femtocells \cite{femto} where multiple LOSs may be created by reflections in close proximity
to the BSs and/or users or may appear in millimeter wave (mmW)
communications \cite{mmv}. With Nakagami-$m$ fading,  stochastic geometry analysis necessitates for tractability
an integer value for $m$ \cite{leila}, thereby limiting the applicability of the model in setup scenarios that capture practical multipath conditions.
Despite the fact that several approaches  show alternative techniques to circumvent such dependency to the Rayleigh fading  \cite{renzo}, \cite{renzo1},\cite{hourani}, \cite{amouri}; there are yet no stochastic geometry models
accounting for state-of-the-art fading models

As a step forward  to bridge this  gap in the literature, this work incorporates versatile multiple-parameter fading models
 into tractable stochastic geometry analysis. These fading models include the  shadowed $\kappa$-$\mu$ distribution, the generalized Rician or the $\kappa$-$\mu$ distribution,  and the $\eta$-$\mu$ distribution.
Besides their elegance, these models are governed by more than two tunable parameters endowing them with high flexibility to capture a broad range of fading
channels, whence their practical significance.   The $\kappa$-$\mu$
distribution, first introduced in \cite{kmu},  can be regarded as a generalization of the classic
Rician fading model for line-of-sight (LOS) scenarios. On the other hand, the
$\eta$-$\mu$ distribution can be considered as a generalization of
the classic Nakagami-$q$ (Hoyt) fading model for non-LOS
scenarios.
Interestingly, the $\kappa$-$\mu$ and the  $\eta$-$\mu$ distributions  represent all-encompassing generalizations, with the classical
channels including  the Nakagami-$m$, the  Hoyt, the Rayleigh, and the Rice fading being their special cases.

The shadowed $\kappa$-$\mu$  fading model, recently introduced in \cite{paris}, jointly includes large-scale
and small-scale propagation effects, by considering that
only the dominant components are affected by shadowing
\cite {paris}.  The shadowed $\kappa$-$\mu$  distribution
includes the shadowed Rician  distribution as special cases, and obviously it also includes the $\kappa$-$\mu$ fading distribution
from which it originates.  However, as we will later see, one of the most appealing properties of the shadowed $\kappa$-$\mu$ distribution
is that it unifies the set of LOS fading models associated with
the $\kappa$-$\mu$ distribution \cite{kmu}, and strikingly, it also unifies the set
of NLOS fading models associated with the $\eta$-$\mu$
distribution \cite{kmu}.  These fading models
offer far  better and much  more flexible representations of practical fading
LOS, NLOS, and shadowed channels than the Rayleigh and Nakagami-$m$ distributions.

Although some works  have considered already shadowed $\kappa$-$\mu$ fading in the context
of stochastic geometry (e.g., \cite{sd}, \cite{d2dsku}), they relied on  series representation
methods (e.g., infinite series in \cite{sd} and Laguerre polynomial series in \cite{d2dsku}) thereby expressing the interference functionals as an infinite series of
higher order derivative terms given by the Laplace transform of the interference power. These methods cannot lend themselves to closed-form expressions and hence require complex numerical evaluation.

To the best of the authors' knowledge, this paper is pioneer in introducing a general approach
of incorporating the comprehensive shadowed $\kappa$-$\mu$, $\kappa$-$\mu$ and   $\eta$-$\mu$ fading models into an exact and unified
stochastic geometry analysis. Besides offering a unified modeling framework for the analysis of a much wider set of practical fading distributions, this work also develops a unified mathematical analysis paradigm for three key performance metrics altogether: i) the average BEP, ii) the coverage probability, and iii) the ergodic achievable rate for cellular networks.\\
The rest of the paper is organized as follows. The system
model, assumptions, and methodology of our new analysis framework are presented
in Section II. In Section III, the baseline downlink modeling
paradigm for  cellular networks over LOS/NLOS and shadowed fading is presented. Our new  unified performance analysis framework is presented in section IV.  Numerical and simulation
results are presented in Section V and the paper is
concluded in Section VI.

\section{ NETWORK AND CHANNEL MODELS}

We consider a downlink single-tier cellular network where single-antennas BSs are deployed
according to a homogeneous PPP $\Psi$ with intensity $\lambda$  and a  typical single-antenna mobile user
 is located at the origin.
It is assumed that all the BSs have the same transmit power $P$. Without loss of generality, all BSs  are assumed to
have an open access policy, and hence, all users can associate with all BSs. The users are assumed to associate to the BSs
according to their  average radio signal strength (RSS)
rule. Similar to \cite[ Sec. VI]{andrews}, universal  frequency reuse is considered with no intra-cell
interference.\\
Further,  we adopt the standard path-loss propagation
model of power attenuation  $r^{-\alpha}$ with the propagation distance $r$, where  $\alpha > 2$ is the path-loss exponent.
For simplicity, we assume that all BSs experience the same path-loss
exponent $\alpha$. Besides we assume that the channel gains between any two  generic locations, denoted by $h$,  include all random channel effects such as fading and shadowing. Additionally, we assume the latter to be
independent of each other, independent of the spatial locations,
symmetric, and identically distributed.
We introduce below some key definitions for the generic channel model distributions adopted in this work.
\subsection{Channel Model Distributions}
\textit{Definition 1}: The shadowed $\kappa$-$\mu$  distribution \cite{paris}.\\
Let $h$ be a random variable statistically following a shadowed $\kappa$-$\mu$
 distribution with mean $\Omega={\cal E}[h]$ and non-negative
real shape parameters $\kappa$, $\mu$, and $m$, i.e., $h\sim S_{\kappa, \mu,m}(\Omega; \kappa,\mu, m)$.
Then its probability density function (PDF) is given by
\begin{equation}
f_{h, S\kappa-\mu}(y)=\frac{ \mu^{\mu} m^{m}(1+\kappa)^{\mu}}{\Gamma(\mu)\Omega^{\mu} (\mu \kappa+m)^{m}}\left(\frac{y}{\Omega}\right)^{\mu-1}e^{-\frac{ \mu(1+\kappa)}{\Omega} y }{\rm {}_{1\!}F_{\!1}}\left(m, \nu , \frac{ \mu^{2} \kappa(1+\kappa)}{\Omega(\mu \kappa+m)} y \right),
\label{pdfsku}
\end{equation}
where ${\rm {}_{1\!}F_{\!1}}(\cdot)$ is the confluent hypergeometric function of \cite[Eq. (13.1.2)]{grad} and the Gamma function is denoted by
$\Gamma(\cdot)$ \cite{grad}.
The shadowed $\kappa$-$\mu$  fading model was originally proposed
in \cite{paris}. In recent works, the shadowed $\kappa$-$\mu$ distribution provided an excellent fit
with channel measurements conducted to characterize the shadowed
fading observed in device-to-device communication
channels \cite{d2d} and in shadowed body-centric communication channels \cite{ban}.
In this new model, the potential clustering of multipath
components is considered alongside the presence of elective
dominant signal components (DSCs), which are subject to Nakagami-$m$
distribution. The shadowed $\kappa$-$\mu$   distribution is an extremely versatile
fading model that also includes as special cases other
important distributions such as the One-Sided Gaussian, Rice, Nakagami-$m$, and Rayleigh distributions.

\textit{Definition 2:} The $\kappa$-$\mu$ distribution \cite{kmu}.\\
Let $h$ be a random variable  statistically following a $\kappa$-$\mu$
distribution with mean  $\Omega={\cal E}[h]$ and non-negative real shape
parameters $\kappa$ and $\mu$, i.e., $h\sim S_{\kappa, \mu}(\Omega; \kappa,\mu)$. Then its pdf is given by
\begin{equation}
f_{h, \kappa-\mu}(y)=\frac{ \mu (1+\kappa)^{\frac{\mu+1}{2}}}{e^{\kappa\mu}\Omega \kappa^{\frac{\mu-1}{2}}}\left(\frac{y}{\Omega}\right)^{\frac{\mu-1}{2}}e^{-\frac{ \mu(1+\kappa)}{\Omega} y }{\rm I}_{\mu-1}
\left(2\mu\sqrt{\frac{ \kappa(1+\kappa)}{\Omega} y }\right),
\label{pdfku}
\end{equation}
where the modified Bessel function of the first kind of order $b$
is represented by ${\rm I}_{b}(\cdot)$ \cite[Eq. 8.431.1]{grad}. The $\kappa$-$\mu$ fading model was originally conceived for modeling the small-scale variations of
a fading signal under line-of-sight (LOS) conditions in non-homogeneous environments. If $\mu=1$, this distribution reduces to the Rice model. The latter
 is the most important
model for representing  a single dominant DSC
between the BS and a mobile user \cite{SEP}. However, multiple DSCs may be created by reflections
from metal objects (e.g., light-posts, cars) in close proximity
to the BSs and/or users, or  may appear in millimeter wave (mmW)
communications where highly directional antennas are used for
short-range communications \cite{mmv}. Therefore, the $\kappa$-$\mu$
distribution offers a much more flexible representation of practical fading
LOS channels than the  Ricean one \cite{beaulieu}.
The $\kappa$-$\mu$ distribution reduces form the shadowed $\kappa$-$\mu$ distribution by eliminating the shadowing of each
dominant component when $m\longrightarrow\infty$.

\textit{Definition 3}: The $\eta$-$\mu$ distribution \cite{kmu}.\\
Let $h$ a random variable statistically following an $\eta$-$\mu$
distribution with mean $\Omega={\cal E}[h]$ and non-negative real shape
parameters $\eta$ and $\mu$, i.e., $h\sim S_{\eta, \mu}(\Omega; \eta,\mu)$. Then its pdf is given
by
\begin{equation}
f_{h, \eta-\mu}(y)=\frac{ \sqrt{\pi}(1+\eta)^{\mu+\frac{1}{2}} \mu^{\mu+\frac{1}{2}}}{\Gamma(\mu)\Omega \sqrt{\eta}(1-\eta)^{\mu-\frac{1}{2}}}\left(\frac{y}{\Omega}\right)^{\mu-\frac{1}{2}}e^{-\frac{ \mu(1+\eta)^{2}}{2\eta\Omega} y }{\rm I}_{\mu-\frac{1}{2}}\left( \frac{ \mu (1-\eta)^{2}}{2 \eta \Omega} y \right).
\label{pdfeu}
\end{equation}
Since it is practically
difficult to achieve a LOS communication all the time,  we consider, in this work,  the $\eta$-$\mu$ distribution as a general fading distribution that can
be used to better represent the small-scale variation of the fading
signal in a NLOS condition \cite{kmu}. If $\mu=m$ and $\eta\longrightarrow\infty$, this distribution reduces to the Nakagami-$m$ model. Moreover, when $\mu=1/2$, we obtain the Nakagami-$q$ Hoyt distribution.

\subsection{Modeling Methodology}
The   instantaneous SINR
for the tagged user placed at the origin\footnote{Based on the properties of homogeneous
PPPs, there is no loss of generality in assuming the tagged user to
be located at the origin \cite{bacelli}.} and located at a random distance
$r$ from its serving BS can be expressed as
\begin{equation}
\text{SINR}=\frac{  P h r^{-\alpha}}{\sigma^{2} + P\sum_{i\in\Psi^{(\setminus 0)}} h_i r_i^{-\alpha}}=\frac{   h r^{-\alpha}}{\frac{\sigma^{2}}{P} + {\cal I} },
\label{SINR}
\end{equation}
where  $\sigma^{2}$ is the noise power, $\Phi^{(\setminus 0)}$ is the point
process representing the interfering BSs (excluding the serving BS) on the tagged channel, and  the random variable ${\cal I}=\sum_{i\in\Psi^{(\setminus 0)}} h_i r_i^{-\alpha}$ denotes  the aggregate interference at the tagged user from $\Psi^{(\setminus 0)}$. According to the
properties of homogeneous PPPs \cite[ Vol. 1, Theorem 1.4.5]{bacelli}, the set of interfering BSs  $ \in \Phi^{(\setminus 0)}$ is still a homogeneous
PPP outside the ball centered at the origin and of radius
$r$. Note that in stochastic geometry analysis,  spatial average performance metrics requires the pdf  of $r$, which is  given  in a PPP
network with RSS association as $f_{r}(x)=2\pi \lambda x e^{-\pi \lambda x^{2}}, \quad r\geq0$ \cite{andrews}.

\textit{Lemma 1:} The MGF of the SINR  can be calculated as
\begin{equation}
M_{\text{SINR}}(s)=1-2{\cal E}_r\left[\sqrt{s  r^{-\alpha}}\int_{0}^{\infty}\underbrace{{\cal E}_{h}\left[\sqrt{h}{\rm J}_1\left(2\sqrt{s h  r^{-\alpha}} \xi\right)\right]}_{\Sigma}e^{-\frac{\sigma^{2}}{P}\xi^{2}}{\cal L}_{{\cal I}}(\xi^{2}) d\xi\right],
\label{mgf}
\end{equation}
where ${\cal E}_x[.]$ is the expectation with respect to the random
variable $x$,  ${\rm J}_{1}(\cdot)$ is the Bessel function of the first kind and first order \cite[Eq. 8.402]{grad}, and  ${\cal L}_{\cal{I}}(s)= {\cal E}\left[e^{-s I}\right]$ denotes the Laplace transform of the aggregate interference.

\textit{Proof:} See Appendix A.

It is worth emphasizing that $\Sigma$ in (\ref{mgf}) is independent of the variable ${\cal L}_{\cal{I}}$
 and is a function of the fading parameters only.
Hence, for  known fading parameters, $\Sigma$ is a constant w.r.t.
the interference Laplace transform. This key  property of Lemma 1 makes the latter a powerful baseline model to build upon in terms of developing tractable analytical models for cellular network,
namely by extending the results  of this paper to many other directions.
Without any pretention of being able to discuss them all due to lack of space,  the most prominent directions  for future works include MIMO and multi-tier downlink performance analysis.
 Although extended in numerous ways to date \cite{andrews}-\!\!\cite{sawy}, these models (i.e., downlink and multi-tier) have never been considered from the standpoint of (\ref{mgf}). Interestingly, Lemma 1  not only promotes general and generic fading channels, but also other generalization aspects such as the the effect of LOS/NLOS propagation where the probability with which a BS is NLOS (also termed
blocking probability) is dependent on the distance between the BS and the receiver of interest \cite{andre1}. In this context, leveraging on Lemma 1,   mathematical models  for millimeter wave
(mmWave) cellular communications,  regarded as a potential scheme in  next fifth generation (5G) systems and Internet of things
(IoT) applications, become tractable.  Remarkably, the proposed framework  is also able to accommodate
 both closest- and strongest-BS association rules as well as single- and multi-slope path loss models \cite{andre2}.
 
Hereafter, by applying (\ref{mgf}),  we characterize the SINR by deriving  its MGF in generalized fading channels.
In contrast to almost all  existing works that adopt the  CCDF-based analysis approach \cite{andrews}-\!\!\cite{leila}, owing to the tractability and favorable
analytical characteristics of Rayleigh fading, we develop  a novel modeling
paradigm for cellular networks that incorporates much more flexible and useful fading models, namely shadowed $\kappa$-$\mu$, $\kappa$-$\mu$, and $\eta$-$\mu$ into a  novel tractable
stochastic geometry  analysis framework. Capitalizing on the several existing MGF-based approaches for performance analysis namely the Gil-Pelaez inversion theorem \cite{gil} for coverage  probability, the  transforms  for rate analysis of K. Hamdi  \cite{hamdi} and  Di Renzo et. al \cite{theo},  Craig's transform for BEP analysis \cite{SEP}, and fractional moment calculation for
error vector magnitude (EVM) performance \cite{evm}, this work develops a unified mathematical paradigm  that bridges the gap between BEP,
coverage probability, and ergodic rate analyses in cellular networks. Extension of this analysis framework to cover performance metrics (EVM, average throughput\footnote{Throughput is defined as the number of successfully transmitted bits per channel use.
}, etc.) under the considered generic fading models
is beyond the scope of this contribution and will be the subject of future works.

\section{UNIFIED ANALYSIS OF THE SINR STATISTICS}
We now state our main and most general results from which all other results in the subsequent sections shall follow.

\textit{Theorem 1:} The  MGF of the SINR over shadowed $\kappa$-$\mu$ fading is
\begin{equation}
M_{\text{SINR}}^{S\kappa\mu}(s)\!=\!1-\frac{\Omega s}{\left(\frac{\mu\kappa}{m}\!+\!1\right)^{m}\!(1\!+\!\kappa)}\!\int_{0}^{\infty} \!\!\!\!{\rm \Psi_1}\!\!\left(\!\mu+1, m; 2,\mu;\frac{- s \xi \Omega}{ \mu(1\!+\!\kappa)}, \frac{\mu \kappa}{\mu \kappa\!+\!m}\!\!\right)\!{\cal E}_{r}\!\! \left[\!\exp\left(\!\!-\xi r^{\alpha}\frac{\sigma^{2}}{P}\!\!\right){\cal L}_{I}^{S\kappa\mu}\left(\xi r^{\alpha} \right)\!\right] d\xi,
\label{MGFSKMU}
\end{equation}
where ${\rm \Psi_1}(\cdot,\cdot;\cdot,\cdot;\cdot,\cdot)$ denotes the Humbert function of the first kind \cite[Eq. 1.2]{humbert}, and  ${\cal L}_{I}^{S\kappa\mu}$ denotes the Laplace transform  of the aggregate
interference when the receiver  interfering link suffers from arbitrary shadowed $\kappa$-$\mu$  fading, i.e., $h_{i\in \Phi^{(\setminus 0)}}\sim S_{\kappa, \mu,m}(\Omega_I; \kappa_I,\mu_I, m_I)$ obtained as
\begin{eqnarray}
{\cal L}_{I}^{S\kappa\mu}(\xi r^{\alpha})= \exp\Bigg(-2 \pi \lambda r^{2}\Bigg( \sum_{k=1}^{m_I}\frac{\binom{k}{m_I}\Xi^{k}\xi^{k}}{\alpha k-2}~{\rm {}_{2}F_{\!1}}\!\left(m_I,k-\frac{2}{\alpha},k+1-\frac{2}{\alpha},-\Xi\xi \right)\nonumber \\ -
\sum_{n=1}^{m_I-\mu_I}\frac{\binom{n}{m_I-\mu_I}\Theta^{n}\xi^{n}}{\alpha n-2}~{\rm {}_{2\!}F_{\!1}}\!\left(m_I,n-\frac{2}{\alpha},n+1-\frac{2}{\alpha},-\Xi \xi\right)\Bigg)\Bigg),
\label{L1}
\end{eqnarray}
when $\mu_I\leq m_I$. And when  $\mu_I \geq m_I$, it becomes
\begin{eqnarray}
{\cal L}_{I}^{S\kappa\mu}(\xi r^{\alpha})&=&\exp\Bigg(-2 \pi \lambda r^{2}\sum_{n,k; (n,k)\neq(0,0)}^{m_I, \mu_I-m_I}\frac{\binom{\mu_I-m_I}{k}\binom{m_I}{n}\Theta^{k} \Xi^{n}\xi^{k+n}}{\alpha (n+k)-2}\nonumber \\ &&{\rm F_{\!1}}\left(n\!+\!k-\!\frac{2}{\alpha},\mu_I\!-\!m_I,m_I, n+k+1-\frac{2}{\alpha},-\Theta\xi, -\Xi\xi \right)\Bigg),
\label{L2}
\end{eqnarray}
where $\Theta=\frac{\Omega_I}{\mu_I(1+\kappa_I)}$ , $\Xi=\frac{ (\mu_I\kappa+m_I)\Omega_I}{m_I\mu_I(1+\kappa_I)}$. Moreover,   ${\rm {}_{2}F_{1}}(a,b,c,x)$  and ${\rm F_{1}}(a,b, b';c ; x , y)$ denote the Gauss hypergeometric function \cite[Eq. 9.100]{grad} and  the first Appell's hypergeometric function \cite[Eq. 9.180.1]{grad}, respectively, and $\binom{a}{b}=\Gamma(a)\Gamma(b)/\Gamma(a+b)$ is the binomial coefficient.\\
\textit{Proof:} See Appendix B.

\textit{Theorem 2:} The MGF of the SINR over $\kappa$-$\mu$ fading is
\begin{equation}
M_{\text{SINR}}^{\kappa\mu}(s)\!\!=\!\!1\!-\!\frac{\Omega e^{-\kappa\mu} s}{(1+\kappa)}\!\int_{0}^{\infty}\!\!\!\! {\rm \Psi_2}\!\left(\mu+1; 2,\mu;\frac{- s \xi \Omega}{\mu (1+\kappa)}, \mu \kappa\right)\!{\cal E}_{r} \left[\exp\left(-\xi r^{\alpha}\frac{\sigma^{2}}{P} \right){\cal L}_{I}^{\kappa\mu}\left(\xi r^{\alpha} \right)\right] d\xi,
\label{MGFKMU}
\end{equation}
where ${\rm \Psi_2}(\cdot,\cdot;\cdot;\cdot,\cdot)$ denotes the Humbert function of the second kind \cite[Eq. 1.3]{humbert}, and  ${\cal L}_{I}^{\kappa\mu}$ denotes the Laplace transform  of the aggregate
interference under $\kappa$-$\mu$ fading, i.e., $h_{i\in \Phi^{(\setminus 0)}}\sim S_{\kappa, \mu}(\Omega_I; \kappa_I,\mu_I)$. Furthermore,  ${\cal L}_{I}^{\kappa\mu}$ is obtained as
\begin{eqnarray}
{\cal L}_{I}^{\kappa\mu}(\xi r^{\alpha})&=&\exp\Bigg(-2 \pi \lambda r^{2} \frac{\mu (1\!+\!\kappa_I)^{\mu_I}}{e^{\kappa_I\mu_I} \Omega^{\mu_I} }\!\!\sum_{k=0}^{\infty}\frac{\left(\frac{\mu_I^{2}\kappa_I(1+\kappa_I)}{\Omega_I}\right)^{k}}{k!}\sum_{n=1}^{\mu_I+k}\frac{\binom{\mu_I+k}{n}\left(\frac{\xi\Omega_I}{\mu_I\kappa_I(1+\kappa_I)}\right)^{n}}{\alpha n-2}\nonumber\\&& {\rm {}_{2}F_{1}}\!\left(\mu_I+k-2,n-\frac{2}{\alpha},n+1-\frac{2}{\alpha},-\frac{\xi\Omega_I}{\mu_I\kappa_I(1+\kappa_I)}\right)\Bigg).
\label{L3}
\end{eqnarray}
\textit{Proof:} See Appendix C.

\textit{Theorem 3:} The MGF of the SINR over $\eta$-$\mu$ fading is
\begin{equation}
M_{\text{SINR}}^{\eta\mu}(s)\!=\!1\!-\!\frac{2 \Omega \eta^{\mu+1} s}{\eta+1}\!\int_{0}^{\infty}\!\!\! \!{\rm \Psi_1}\!\!\left(\!2\mu+1, \mu; 2,2\mu;\frac{- s \xi \Omega \eta}{ \mu (1+\eta)}, 1-\eta\!\right){\cal E}_{r} \left[\!\exp\left(\!-\xi r^{\alpha}\frac{\sigma^{2}}{P}\! \right){\cal L}_{I}^{\eta\mu}\left(\xi r^{\alpha} \!\right)\!\right]d\xi,
\label{mgfkmu}
\end{equation}
where the Laplace transform  of the aggregate
interference $h_{i\in \Phi^{(\setminus 0)}}\sim S_{\eta, \mu}(\Omega_I; \eta_I,\mu_I)$, denoted as ${\cal L}_{I}^{\eta \mu}$,  is obtained as
\begin{eqnarray}
{\cal L}_{I}^{\eta\mu}(\xi r^{\alpha})&=&\exp\Bigg(\!\!-2 \pi \lambda r^{2}\sum_{n,k; (n,k)\neq(0,0)}^{\mu_I, \mu_I}
\frac{\binom{\mu_I}{k}\binom{\mu_I}{n}\eta_I^{-k}\left(\frac{\Omega_I\eta_I \xi}{\mu_I(1+\!\eta_I)}\right)^{n+k}}{\alpha(n+ k)\!-\!2}\nonumber \\&&{\rm F_{\!1}}\left(\!n\!+\!k\!-\!\frac{2}{\alpha},\mu_I,\mu_I, n\!+k+1-\frac{2}{\alpha},-\frac{\Omega_I\xi}{\mu_I(1+\eta_I)}, -\frac{\Omega_I\eta_I\xi}{\mu_I(1\!+\!\eta_I)} \!\!\right)\!\!\Bigg).
\label{L4}
\end{eqnarray}
\textit{Proof:} From (\ref{pdfsku}), when $m=\mu/2$,  we resort to the
reduction formula of  ${\rm {}_{1\!}F_{\!1}}(\cdot)$ given by \cite[Eq. 9.6.47]{grad}
\begin{equation}
{\rm {}_{1}F_{1}}\left(\beta, 2 \beta, z\right)=2^{2\beta-1}\Gamma\left(\beta+\frac{1}{2}\right)z^{\frac{1}{2}-\beta}e^{z/2}{\rm I}_{\beta-\frac{1}{2}}\left(\frac{z}{2}\right),
\end{equation}
readily yielding (\ref{pdfeu}) after some  algebraic manipulations.   Previously shown in  \cite{unif}, this result reveals that the $\eta$-$\mu$ fading distribution arises
as a particular case of the more general  shadowed $\kappa$-$\mu$ model.  Notice that, the Nakagami-$q$ (Hoyt) model with shape parameter $q =\frac{1}{\sqrt{2 \kappa+1}}$  arises when
$m = \frac{\mu}{2}= 0.5$, since $\eta = q^{2}$ for the $\eta$-$\mu$ distribution in (\ref{pdfeu})\cite{unif}\footnote{The $\eta$-$\mu$ fading model is symmetrical for $\eta$ $\in$ $[0,1]$ and $\eta$ $\in$ $[1,\infty]$.}.
Accordingly, the SINR MGF under $\eta$-$\mu$ fading is obtained from  (\ref{MGFSKMU}) by setting $\underline{m}=\mu$, $\underline{\mu}=2\mu$, and $\underline{\kappa}=\frac{1-\eta}{2 \eta}$. Since $m_I=\mu_I/2\leq\mu_I$, then ${\cal L}_{I}^{\kappa\mu}$ reduces from ${\cal L}_{I}^{S\kappa\mu}$ in (\ref{L2}) by setting ${\underline m}_I=\mu_I$, ${\underline \mu}_I=2\mu_I$, and ${\underline \kappa}_I=\frac{1-\eta_I}{2 \eta_I}$, thereby yielding (\ref{L4}).

\textit{Corollary 1: } The MGF of the SINR over arbitrary Nakagami-$m$ fading is given by
\begin{equation}
M_{\text{SINR}}^{m}(s)=1-\Omega s\int_{0}^{\infty} {\rm {}_{1\!}F_{\!1}}\left(m+1, 2;\frac{- s  \Omega}{ m}\xi\right){\cal E}_{r} \left[\exp\left(-\xi r^{\alpha}\frac{\sigma^{2}}{P} \right){\cal L}_{I}^{m}\left(\xi r^{\alpha} \right)\right] d\xi,
\label{mgfnak}
\end{equation}
and  the Laplace transform of the aggregate interference under Nakagami-$m$  ${\cal L}_{I}^{m}$  is given by
 \begin{equation}
{\cal L}_{I}^{m}(\xi r^{\alpha})=\exp\left(-\pi \lambda r^{2} \left({\rm {}_{2\!}F_{\!1}}\left(-\frac{2}{\alpha}, m_I, 1-\frac{2}{\alpha}; -\frac{\Omega_I}{m_I}
\xi\right)-1\right)\right).
\label{lm}
\end{equation}
\textit{Proof:}  The Nakagami-$m$ fading distribution arises
as a particular case of the more general shadowed  $\kappa$-$\mu$ model  when $m=\mu$. However, this simplification is not straightforward and actually requires further involved manipulations given in Appendix D.

It is worthwhile to note that ${\cal L}_{I}^{m}$ in (\ref{lm}) is a well-known result in the area of cellular networks analysis over Nakagami-$m$ fading \cite{MIMO2},\!\cite{leila}. While ${\cal L}_{I}^{m}$ has so far been presented  as a fundamental finding in previous works, it becomes in this contribution  a secondary result that simply reduces from a more general performance analysis framework.

The MGF of the SINR for Rayleigh  fading,   extensively adopted in the literature \cite{andrews}-\!\!\cite{sawy},   reduces simply from (\ref{mgfnak}) when $m=m_I=1$  as\vspace{-0.1cm}
\begin{eqnarray}
M_{\text{SINR}}(s)=1-\Omega s\int_{0}^{\infty} \exp(- s \Omega \xi){\cal E}_{r}\left[\exp\left(-\xi r^{\alpha}\frac{\sigma^{2}}{P} \right){\cal L}_{I}\left(\xi r^{\alpha} \right)\right] d\xi,
\label{mgfray}
\end{eqnarray}
where    the Laplace transform of the aggregate interference under Rayleigh fading, ${\cal L}_{I}$, is  a special case of (\ref{lm}) when  $m_I=1$.

\textit{Proof:} This proof is a special case of \textit{Theorem 4} with more simplifications arising from the fact that $ {\rm {}_{1}F_{1}}\left(a, a;x\right)=\exp(x)$. \\
For completeness, it is worthwhile to mention that (\ref{mgfray}) can be easily deduced from CCDF-based analysis frameworks \cite{andrews}-\!\cite{sawy}  by applying $M(s)=1-s {\cal L}_{\mathbb{P}(\text{SINR}>T)}(s)$  where $\mathbb{P}(\text{SINR}>T)$ is the CCDF of the SINR, given  for instance in \cite{andrews} as $\mathbb{P}(\text{SINR}>T)={\cal E}_{r}\left[\exp(-\frac{T}{P \Omega} r^{\alpha}\sigma^{2} ){\cal L}_{I}\left(\frac{T}{\Omega} r^{\alpha} \right)\right]$, and carrying the change of variable $\xi=T/\Omega$. This key observation, unambiguously, corroborates the much wider scope claimed by our novel  analysis framework and the rigor of its mathematical derivations.

While applying the shadowed $\kappa$-$\mu$ or  the $\kappa$-$\mu$ (to capture different DSCs scenarios) to the tagged user link is quite
intuitive (typically in the case of  future femtocells and picocells), it might not be as much obvious to do so  to the interference links. Actually, in a typical urban deployment,  interfering channels are less likely to experience LOS than the direct link. However,  destructive LOS interference may also happen in practice, namely in suburban and rural areas having wide parks and open spaces.  In this work, the treatment of the tagged user  link  is independent from its interference counterpart as can be seen from (\ref{mgf}). This dissociation is very appreciable since it allows modeling cellular networks with direct and interfering links experiencing asymmetric fading (i.e., different fading models). Although not shown  explicitly in this work, cellular networks performance under asymmetric fading can be easily assessed by swapping ${\cal L}_{I}$ in (\ref{L1}), (\ref{L2}), (\ref{L3}), (\ref{L4}), and (\ref{lm}).

The new fundamental statistics disclosed in Theorems 1 to 4
provide a novel unifying analysis framework for of a variety of extremely important fading distributions. In some particular cases, the obtained formulas reduce to previously well-known major results in the literature.
Besides, even though this work focuses on the shadowed $\kappa$-$\mu$, $\kappa$-$\mu$, and $\eta$-$\mu$ distributions, our new analysis framework is extensible to  any other  fading/shadowing distribution as long as the quantity pertaining to the expectation over $h$ in (\ref{mgf}) (i.e., $\Sigma$) can be obtained in closed form.  Moreover,  by assuming composite fading/shadowing
fading over the interfering links, the analytical tractability of our new analysis framework is not affected at all and we can still formulate all performance metrics
with the Laplace transform of the aggregate interference. Nevertheless, the expression of the latter might may become
more involved.

\section{AVERAGE ACHIEVABLE RATE}
The average transmission rate, as defined by Shannon's capacity, is evaluated using the MGF transform in \cite[Lemma 21]{hamdi} as\vspace{-0.2cm}
\begin{equation}
C\triangleq{\cal E} [\ln\left(1+\text{SINR}\right)]=\int_{0}^{\infty}\!\!\frac{\exp\{-s\}}{s}\left(1-M_{\text{SINR}}(s)\right)ds.
\label{C}
\end{equation}
The average rate is computed in  nats/Hz ($1$ bit $= \ln(2) = 0.6934$ \text{nats}) for a typical
user assumed to achieve the Shannon bound at its instantaneous SINR.
We state now the main theorems that give the
ergodic capacity of a typical mobile user on the downlink.

\textit{Theorem 5: } The average ergodic rate of a typical mobile
user on the downlink  over shadowed $\kappa$-$\mu$ fading is\vspace{-0.1cm}
\begin{equation}
C^{S\kappa\mu}(\lambda, \alpha )\!\!=\!\!\frac{\Omega }{\left(\frac{\mu\kappa}{m}\!+\!1\right)^{m}(1\!+\!\kappa)}\int_{0}^{\infty}\!\!\!\!{\rm F_2}\!\left(\!\mu\!+\!1, m, 1; 2,\mu;\frac{\mu \kappa}{\mu \kappa\!+\!m},\frac{-  \xi \Omega}{ \mu(1\!+\!\kappa)} \!\right)\!{\cal E}_{r}\! \left[\exp\!\left(\!-\xi r^{\alpha}\frac{\sigma^{2}}{P}\! \right)\!{\cal L}_{I}^{S\kappa\mu}\left(\!\xi r^{\alpha} \!\right)\!\right] d\xi,
\label{C1}
\end{equation}
where ${\rm F_{2}}(a,b, b';c , c'; x , y)$ stands for the Appell's hypergeometric function of the second kind \cite[Eq. 9.180.2 ]{grad}, and   ${\cal L}_{I}\left(\xi r^{\alpha} \right)$ is given in (\ref{L1})-(\ref{L2}).\\
\textit{Proof:}
Plugging (\ref{MGFSKMU}) into (\ref{C}), and resorting to
\begin{equation}
{\rm F_2}\left(a, b, b'; c,c';x,y\right)=\frac{1}{\Gamma(b')}\int_{0}^{\infty}t^{b'-1}e^{-t}{\rm \Psi_1}\left(a, b; c, c'; x, yt\right) dt,
\end{equation}
yields the desired result after some manipulations.

A hallmark of current-small cell systems in urban environments
is that they are overwhelmingly interference-limited, where the downlink channels are severed by interference rather than by thermal noise,
especially at  the cell edge where the interference power is typically
so much larger. In such a case,  the average rate is limited by bandwidth rather than power. Therefore, the case of no noise (or infinite
transmit power $P$) is of particular interest because it
captures the scenario where the transmit power would not be a binding
constraint over downlink communications.

\textit{Corollary 1 (shadowed $\kappa$-$\mu$, no noise or infinite $P$): }
 The interference-limited ergodic rate of a typical mobile
user on  the downlink  over shadowed $\kappa$-$\mu$ fading is
\begin{equation}
C^{S\kappa\mu, \infty}( \alpha )=\frac{\Omega }{\left(\frac{\mu\kappa}{m}+1\right)^{m}(1+\kappa)}\int_{0}^{\infty}\frac{{\rm F_2}\left(\mu+1, m, 1; 2,\mu;\frac{\mu \kappa}{\mu \kappa+m},\frac{-  \xi \Omega}{ \mu(1+\kappa)} \right)}{1+{\cal A}(\xi)} d\xi,
\label{CN1}
\end{equation}
where
\begin{equation}
{\cal A}(\xi)\!\!=\!\!\left\{
                \begin{array}{ll}
                  \!\!\!\sum_{k=1}^{m_I}\!\!\frac{\binom{m_I}{k}\Xi^{k}\xi^{k}}{\frac{\alpha k}{2}-1}{\rm {}_{2}F_{1}}\!\!\left(\!m_I,k\!-\!\frac{2}{\alpha},k\!+\!1\!-\!\frac{2}{\alpha},-\Xi\xi \!\right)\!-\!\!
\sum_{n=1}^{m_I\!-\!\mu_I}\!\!\frac{\binom{m_I\!-\!\mu_I}{n}\Theta^{n}\xi^{n}}{\frac{\alpha n}{2}\!-\!1}{\rm {}_{2}F_{\!1}}\!\!\left(m_I,n\!-\!\frac{2}{\alpha},n\!+\!1\!-\!\frac{2}{\alpha},-\Xi \xi\right),  \\ \quad\quad\quad\quad\quad\quad\quad\quad\quad\quad\quad\quad\quad\quad\quad\quad\quad\quad\quad\quad\quad\quad\quad\quad\quad\quad\quad\quad\quad\quad\quad\quad\quad\quad\hbox{$\mu_I\leq m_I$;} \\
            \!\!\! \sum_{n,k; (n,k)\neq(0,0)}^{m_I, \mu_I-m_I}\!\!\!\frac{\binom{\mu_I-m_I}{k}\binom{m_I}{n}\Theta^{k} \Xi^{n}\xi^{k+n}}{\frac{\alpha(n+k)}{2}-1}{\rm F_{\!1}}\left(\!n\!+\!k\!-\!\frac{2}{\alpha},\mu_I\!-\!m_I,m_I, n\!+\!k\!+\!1\!-\!\frac{2}{\alpha},-\Theta\xi, -\Xi\xi \right), \hbox{$\mu_I> m_I$.}
                \end{array}
              \right.
\end{equation}
\textit{Proof:}
When noise is neglected, i.e.,  $\sigma^{2}\simeq 0$ or the  transmit power is not a binding
constraint i.e., $P\simeq\infty$, then the expectation ${\cal E}_{r} \left[\exp\left(-\xi r^{\alpha}\frac{\sigma^{2}}{P} \right){\cal L}_{I}^{S\kappa\mu}\left(\xi r^{\alpha} \right)\right]\underset{\sigma^{2}\simeq 0, P\simeq\infty}{\approx}{\cal E}_{r} \left[{\cal L}_{I}\left(\xi r^{\alpha} \right)\right]$  with respect to the distance $r$ separating a typical
user from its tagged BS with pdf $f_{r}(x)=2\pi \lambda x e^{-\pi \lambda x^{2}}$  is expressed in closed form using \cite[Eq. 3.326.2]{grad}, thereby yielding the simplified expressions of the ergodic rate shown in (\ref{CN1}).

\textit{Theorem 6:} The average  rate of a typical mobile
user at a distance $r$ from its serving BS over $\kappa$-$\mu$ fading is
\begin{equation}
C^{\kappa\mu}(\lambda, \alpha )=\frac{\Omega e^{-\kappa\mu} }{(1+\kappa)}\int_{0}^{\infty}{\rm \Psi_1}\left(\mu+1, 1; 2,\mu; \mu \kappa ,\frac{-  \xi \Omega}{\mu (1+\kappa)}\right){\cal E}_{r} \left[\exp\left(-\xi r^{\alpha}\frac{\sigma^{2}}{P} \right){\cal L}_{I}^{\kappa\mu}\left(\xi r^{\alpha} \right)\right] d\xi,
\label{C2}
\end{equation}
where ${\cal L}_{I}\left(\xi r^{\alpha} \right)$ is given in (\ref{L3}).\\
\textit{Proof:} The results follows  after employing substituting (\ref{MGFKMU}) into (\ref{C}) and  applying\vspace{-0.2cm}
\begin{equation}
{\rm \Psi_1}\left(a, b; c,c';x,y\right)=\frac{1}{\Gamma(b)}\int_{0}^{\infty}t^{b-1}e^{-t}{\rm \Psi_2}\left(a, c, c'; x, yt\right) dt.
\end{equation}
Another rationale to get (\ref{C2}) starts from (\ref{C1}) and employs the following limit relation \cite{humbert}:
\begin{equation}
\lim_{\epsilon\rightarrow0} {\rm F_{2}}\left(\alpha,\frac{b'}{\epsilon},b,c',c;\epsilon x, y\right)={\rm \Psi_1}\left(\alpha, b; c',c; b'x , y\right).
\end{equation}

\textit{Corollary 2 ($\kappa$-$\mu$, no noise or infinite $P$): }
The interference-limited ergodic rate of a typical mobile
user and its serving BS over  $\kappa$-$\mu$ fading is
\begin{equation}
C^{\kappa\mu, \infty}( \alpha )\!\!=\!\!\frac{\Omega e^{-\kappa\mu}}{(1+\kappa)}\!\!\int_{0}^{\infty}\!\!\!\!\!\!\! \frac{{\rm \Psi_1}\left(\mu+1, 1; 2,\mu; \mu \kappa ,\frac{-  \xi \Omega}{\mu (1+\kappa)}\right)d\xi}{1\!+\! \frac{\mu_I (1\!+\!\kappa_I)^{\mu_I}}{e^{\kappa_I\mu_I} \Omega_I^{\mu_I} }
\sum_{k=0}^{\infty}\!\!\frac{\left(\frac{\mu_I^{2}\kappa_I(1\!+\!\kappa_I)}{\Omega_I}\right)^{k}}{k!}\sum_{n=1}^{\mu_I\!+\!k}\!\!\frac{\binom{\mu_I+k}{n}\left(\frac{\xi\Omega_I}{\mu_I\kappa_I(1\!+\!\kappa_I)}\right)^{n}{\rm {}_{2}F_{1}}\left(\!\mu_I\!+\!k\!-\!2,n\!-\!\frac{2}{\alpha},n\!+\!1\!-\!\frac{2}{\alpha},-\frac{\xi\Omega_I}{\mu_I\kappa_I(1\!+\!\kappa_I)}\!\right)}{\frac{\alpha n}{2}-1}} .
\label{Ckuh}
\end{equation}
\textit{Proof:} The result follows from (\ref{C2}) after setting $\sigma^{2}=0$ and  using \cite[Eq. 3.326.2]{grad}.

\textit{Corollary 3  (Rice fading):} An interesting case to be addressed here is the typical Rice model, which arises from the $\kappa$-$\mu$ fading  when $\mu=1$ and $\kappa=K$ where $K$ is the Rice factor.
The ergodic rate under Rice fading  is obtained from (\ref{C2}) as
\begin{equation}
C^{Rice}(\lambda, \alpha )=\frac{\Omega e^{-K} }{1+K}\int_{0}^{\infty}{\rm \Psi_1}\left(2, 1; 2,1; K ,\frac{-  \xi \Omega}{1+K}\right){\cal E}_{r} \left[\exp\left(-\xi r^{\alpha}\frac{\sigma^{2}}{P} \right){\cal L}_{I}^{Rice}\left(\xi r^{\alpha} \right)\right] d\xi,
\label{Crice}
\end{equation}
where\vspace{-0.3cm}
\begin{equation}
{\cal L}_{I}^{Rice}(\xi r^{\alpha})\!\!=\!\!\exp\!\!\Bigg(\!\!-2 \pi \lambda  \frac{r^{2}e^{-K}}{\alpha\Omega } \!\!\sum_{k=0}^{\infty}\frac{\left(\frac{K(1\!+\!K)}{\Omega}\right)^{k}}{k! }\!\!\sum_{n=1}^{k+1}\frac{\binom{k+1}{n}\left(\frac{\xi\Omega}{K(1\!+\!K)}\right)^{n}}{n\!-\!\frac{2}{\alpha}} \!\!{\rm {}_{2}F_{1}}\!\left(\!k\!-\!1,n\!-\!\frac{2}{\alpha},n\!+\!1\!-\!\frac{2}{\alpha},-\frac{\xi\Omega}{K(1\!+\!K)}\!\!\right)\!\!\Bigg).
\label{Lrice}
\end{equation}
\textit{Theorem 7:} The average  rate of a typical mobile
user at a distance $r$ from its serving BS  over $\eta$-$\mu$ fading is
\begin{equation}
C^{\eta\mu}(\lambda, \alpha )\!\!=\!\!\frac{2 \Omega \eta^{\mu+1} }{\eta+1}\!\!\int_{0}^{\infty}\!\!{\rm F_2}\left(2\mu+1, \mu, 1; 2,\mu;\frac{- \xi \Omega \eta}{ \mu (1+\eta)}, 1-\eta\right){\cal E}_{r} \left[\exp\left(-\xi r^{\alpha}\frac{\sigma^{2}}{P} \right){\cal L}_{I}^{\eta\mu}\left(\xi r^{\alpha} \right)\right] d\xi,
\label{Cemu}
\end{equation}
where ${\cal L}_{I}\left(\xi r^{\alpha} \right)$ is given in (\ref{L3}).\\
\textit{Proof:}  The result is obtained along the same lines of (\ref{mgfkmu}) by performing similar substitutions in (\ref{CN1}). Moreover, ${\cal L}_{I}^{\eta\mu}\left(\xi r^{\alpha} \right)$ is given in (\ref{L3}).

 \textit{Corollary 4 ($\eta$-$\mu$,  no noise or infinite $P$):}
The interference-limited ergodic rate of a typical mobile
user on the downlink over $\eta$-$\mu$ fading is
\begin{equation}
C^{\eta\mu, \infty}(\alpha )=\int_{0}^{\infty}\!\!\!\!\frac{\frac{2 \Omega \eta^{\mu+1} }{\eta+1}{\rm F_2}\left(2\mu\!+\!1, \mu, 1; 2,2\mu;\frac{- \xi \Omega \eta}{ \mu (1+\eta)}, 1-\eta\right)d\xi}{1\!\!+ \!\!\!\sum_{\overset{n,k}{ (n,k)\neq(0,0)}}^{\mu_I, \mu_I}\!\!\!\!\!\!\!\!\frac{\binom{\mu_I}{k}\binom{\mu_I}{n}\eta_I^{-k}\left(\frac{\Omega_I\eta_I\xi}{\mu_I(1+\eta_I)}\right)^{n+k}}{\frac{\alpha (n+ k)}{2}-1}{\rm F_{1}}\!\left(\!n\!+\!k\!-\!\frac{2}{\alpha},\mu_I,\mu_I, n\!+\!k\!+\!1\!-\!\frac{2}{\alpha},-\frac{\Omega_I\xi}{\mu_I(1\!+\!\eta_I)}, -\frac{\Omega_I\eta_I\xi}{\mu_I(1\!+\!\eta_I)} \right)} .
\label{Cmeuh}
\end{equation}
\textit{Proof:} Substituting (\ref{L3}) into (\ref{Cemu}) with $\sigma^{2}=0$ and using \cite[Eq. 3.326.2]{grad} yield the desired result.

\textit{Theorem 8:}
 The average ergodic rate of a typical mobile
user  over Nakagami-$m$ fading is
\begin{eqnarray}
C^{m}(\lambda, \alpha )=\int_{0}^{\infty}\frac{1}{\xi}\left(1-\left(1+\frac{\Omega \xi}{m}\right)^{-m}\right){\cal E}_{r} \left[\exp\left(-\xi r^{\alpha}\frac{\sigma^{2}}{P} \right){\cal L}_{I}^{m}\left(\xi r^{\alpha} \right)\right] d\xi,
\label{Cnak}
\end{eqnarray}
where ${\cal L}_{I}^{m}\left(\xi r^{\alpha} \right)$ is given in (\ref{lm}).\\
\textit{Proof:} The result is a special case of (\ref{CN1}) when  $m=\mu$. In this case, a reduction formula of the Appell's ${\rm F_{2}}$ function is given in Appendix E. Alternatively, one can obtain (\ref{Cnak}) after plugging (\ref{mgfnak}) into (\ref{C}) and resorting to \cite[Eq. 7621.5]{grad}.

 \textit{Corollary 5 (Nakagami-$m$, no noise or infinite $P$):}
The interference-limited ergodic rate of a typical mobile
user on  the downlink  over Nakagami-$m$ fading is obtained as
 \begin{equation}
C^{m, \infty}(\alpha )=\int_{0}^{\infty}\frac{1-\left(1+\frac{\Omega \xi}{m}\right)^{-m}}{\xi{\rm {}_{2}F_{1}}\left(\frac{-2}{\alpha}, m_I, 1-\frac{2}{\alpha}; -\frac{\Omega_I}{m_I}\xi\right)} d\xi.
\label{cmh}
\end{equation}
The Rayleigh case reduces from (\ref{cmh}) when $m=m_I=1$; a key result previously obtained in \cite[Theorem 3]{andrews},  under, however, a more involved expression that encompasses a two-fold integration.

Equations (\ref{CN1}), (\ref{Ckuh}), (\ref{Cmeuh}), and (\ref{cmh}) show that if $\sigma^{2}\rightarrow 0$ (or the transmit power is
not a binding constraint, i.e., $P=\infty$), the  average rate in a single-tier cellular network is independent
of the BS intensity. This result is in compliance with those disclosed
in \cite{andrews}. In terms of
average spectral efficiency, the downlink SINR performance (in case of non-binding maximum transmit power constraint) is independent of the BS
intensity. Therefore, interference
management techniques such as  frequency reuse,
interference cancellation, MIMO communications,  interference avoidance, inter-cell cooperation, etc.,  are indeed overriding in order to increase the average rate in interference-limited cellular networks.

\section{COVERAGE PROBABILITY}
The cellular coverage probability is
defined as
\begin{equation}
P_{cov}(T)\triangleq \mathbb{P}\left(\text{SINR}\geq T\right),
\end{equation}
where $T$ represents the minimum SINR value for reliable downlink connection.

\textit{Remark: Laplace transform inversion.}
The Laplace transform of the complementary cumulative distribution function (CCDF) of the SINR, that
is $ P_{cov}(T)=\mathbb{P}\left(\text{SINR}\geq T\right)$, is related to the Laplace
transform of the SINR as follows
\begin{eqnarray}
{\cal L}_{P_{cov}}(z)&=&\int_{0}^{\infty}P_{cov}(y)e^{-z y}dy \nonumber \\
&\overset{(a)}{=}&\left[P_{cov}(y)\frac{-e^{-z y}}{z}\right]_{0}^{\infty}-\int_{0}^{\infty}P'_{cov}(y)\frac{-e^{-z y}}{z}dy\nonumber \\
&\overset{(b)}{=}&\frac{1-{\cal E}[e^{-z \text{SINR}}]}{z}=\frac{1}{z}-\frac{M_{\text{SINR}}(z)}{z}, \quad  z\in R_{+}
\label{cdf1}
\end{eqnarray}
where equality $(a)$ is due to integration by parts and equality $(b)$ follows form the definition of the MGF. The SINR CCDF
 $ P_{cov}(T)$  may be retrieved from its Laplace transform using the Euler characterization as
\cite[Eq. 2]{abat}
\begin{equation}
P_{cov}(T)\triangleq \mathbb{P}\left(\text{SINR}\geq T\right)=\frac{2 e^{b T }}{\pi}\int_{0}^{\infty}\text{Re}\left[{\cal L}_{P_{cov} }(b+i u)\right]\cos( u T)du, \quad T\geq0,
\label{cdf2}
\end{equation}
where $i^{2}=-1$ and $b>0$ is such that ${\cal L}_{P_{cov}}$ has no singularities on or to the
right of it. The above inversion may be carried numerically
using the Abate and Whitt algorithm \cite[Eq. 15]{abat}
\begin{equation}
P_{cov}(T)\simeq\frac{2^{-m} e^{A/2 }}{T}\sum_{k=0}^{m}\binom{m}{k}\sum_{l=0}^{n+k}\frac{(-1)^{l}\text{Re}\left[{\cal L}_{P_{cov} }\left(\frac{A+2i\pi l}{2 T}\right)\right]}{l+\textbf{1}\{l=0\}}, \quad T\geq0,
\label{num}
\end{equation}
with a typical choice of $A=18.4$, $m=11$, and $n=15$.

The following propositions provide the downlink SINR in cellular networks over the considered fading models.

\textit{Proposition 1: } The  coverage probability   of a typical-randomly
located mobile user in the general cellular network
model of Section II over shadowed $\kappa$-$\mu$ fading is
\begin{eqnarray}
P_{cov}^{S\kappa\mu}(T)\!\!\!&=&\!\!\!\frac{2 e^{b T }\Omega}{\pi \left(\frac{\mu\kappa}{m}+1\right)^{m}(1+\kappa)}\int_{0}^{\infty}\!\!\! \int_{0}^{\infty}\!\!\!\text{Re}\left[{\rm \Psi_1}\left(\mu+1, m; 2,\mu;\frac{- (b+i u) \xi \Omega}{ \mu(1+\kappa)}, \frac{\mu \kappa}{\mu \kappa+m}\right)\right]\cos( u T )du\nonumber\\  && {\cal E}_{r} \left[\exp\left(-\xi r^{\alpha}\frac{\sigma^{2}}{P} \right){\cal L}_{I}^{S\kappa\mu}\left(\xi r^{\alpha} \right)\right] d\xi.
\label{pcovsku}
\end{eqnarray}
\textit{Proof:} Combining (\ref{cdf1}) and (\ref{cdf2})  and using (\ref{MGFSKMU}) yield the result after some manipulations.

\textit{Proposition 2: } The coverage probability   of a typical randomly-located mobile user in the general cellular network
model of Section II over  $\kappa$-$\mu$ fading is
\begin{eqnarray}
P_{cov}^{\kappa\mu}(T)&=&\frac{2 e^{b T }\Omega e^{-\kappa\mu}}{\pi(1+\kappa)}\int_{0}^{\infty}\!\!\! \int_{0}^{\infty}\!\!\!\text{Re}\left[{\rm \Psi_2}\left(\mu+1; 2,\mu;\frac{- (b+i u) \xi \Omega}{\mu (1+\kappa)}, \mu \kappa\right)\right]\cos( u T )du \nonumber\\  && {\cal E}_{r} \left[\exp\left(-\xi r^{\alpha}\frac{\sigma^{2}}{P} \right){\cal L}_{I}^{\kappa\mu}\left(\xi r^{\alpha} \right)\right] d\xi.
\label{pckmu}
\end{eqnarray}
\textit{Proof:} The result is obtained along the same lines of (\ref{pcovsku}) with the difference of using (\ref{MGFKMU}).

As for $\eta$-$\mu$ fading, the result follows along the same lines of (\ref{pckmu}) with the use of the appropriate SINR's MGF expression given in (\ref{mgfkmu}). The result is omitted here for the sake of conciseness.

\textit{Proposition 3: }  In a Rician fading environment with parameters $K$ and $\Omega$, the coverage probability is given by
\begin{equation}
P_{cov}^{Rice}(T)\!\!=\!\!\frac{2 e^{b T }e^{-K }\Omega }{\pi (1\!+\!K)}\!\!\int_{0}^{\infty}\!\! \left(\!\int_{0}^{\infty}\!\!\!\!\!\text{Re}\!\!\left[\!{\rm \Psi_2}\!\!\left(\!2; 2,1;\frac{- (b\!+\!i u) \xi \Omega}{ 1\!+\!K}, K\!\right)\!\right]\!\!\cos(u T)du\!\!\right)\!{\cal E}_{r}\!\! \left[\!\exp\left(\!-\xi r^{\alpha}\frac{\sigma^{2}}{P} \right)\!{\cal L}_{I}^{Rice}\left(\xi r^{\alpha} \!\right)\!\right]d\xi.
\end{equation}
\textit{Proof:} The coverage probability under Rician fading  is derived from (\ref{pckmu}) by setting $\mu=1$ and $\kappa=K$.

\textit{Proposition 4: }  In Nakagami-$m$, the coverage probability is given by
\begin{equation}
p_{cov}^{m}(T)=\frac{2 e^{b T }\Omega }{m}\!\!\int_{0}^{\infty}\!\!\! \left(\int_{0}^{\infty}\!\!\!\!\text{Re}\!\!\left[\!{\rm {}_{1\!}F_{\!1}}\!\left(\!m\!+\!1, 2;\frac{- (b+i u) \Omega}{ m}\xi\!\right)\!\right]\cos(u T)du\!\right)\!{\cal E}_{r} \!\left[\!\exp\left(-\xi r^{\alpha}\frac{\sigma^{2}}{P}\! \right){\cal L}_{I}^{m}\left(\xi r^{\alpha}\! \right)\!\right]d\xi.
\label{pnak}
\end{equation}
\textit{Proof:} Combining (\ref{cdf1}) and (\ref{cdf2})  and using (\ref{mgfnak}) yield the desired result after some manipulations. Note that (\ref{pnak}) could also be derived from (\ref{pcovsku})  by  setting $\mu=m$ and form  (\ref{pckmu}) when $\kappa\rightarrow0$ \cite{paris}.

Under the commonly-used Rayleigh fading, the coverage probability follows from substituting (\ref{mgfray}) into (\ref{cdf2}). Resorting to the fact that
\begin{equation}
\int_{0}^{\infty}\text{Re}\left[\exp(- (b+i u) \xi \Omega)\right]\cos( u T )du=\frac{e^{-b \xi \Omega}}{2}\delta[T-\Omega \xi],
\end{equation}
where $\delta(\cdot)$ is the Dirac delta function, it follows that under Rayleigh fading,  $P_{cov}$ is given by
\begin{equation}
p_{cov}^{R}(T)=\Omega e^{b T }\!\! \int_{0}^{\infty}\!\!\!\!\!\!e^{-b \xi \Omega} \delta[T-\Omega \xi]{\cal E}_{r} \left[\exp(-\xi r^{\alpha}\frac{\sigma^{2}}{P} ){\cal L}_{I}\left(\xi r^{\alpha} \right)\right]d\xi={\cal E}_{r} \left[\exp(-\frac{T }{\Omega} r^{\alpha}\sigma^{2} ){\cal L}_{I}\left(\frac{T}{ \Omega} r^{\alpha} \right)\right].
\label{pray}
\end{equation}
The last expression in (\ref{pray}) matches the well-known main result for Rayleigh fading in \cite{andrews}, validating once again the wider scope of our new analysis approach.

So far, we have been able to provide a unified framework for cellular networks performance through the derivation of average rate and coverage probability, two metrics that are agnostic towards the modulation scheme and the receiver
type. However, to capture more system details, we aim in the following to extend our new stochastic geometry
analysis framework to a tangible error performance metric, namely the BEP.

\section{AVERAGE BEP UNDER GAUSSIAN SIGNALING APPROXIMATION}
This section delves into fine wireless communication details trough BEP analysis.
In the context of wireless networks, error probability performance
has mainly been studied and conducted over additive
white Gaussian noise (AWGN) or Gaussian-interference channels
\cite{SEP}.  Without loss of generality, we focus on the BEP, denoted by ${\cal B}$, for coherent M-PSK (phase shift keying) and M-QAM (quadrature
amplitude modulation) constellations  given by \cite{SEP},~\cite{pro} as
\begin{equation}
{\cal B}=\beta_M \sum_{p=1}^{\tau_M}{\cal E}\left[{\rm Q}\left(a_p \sqrt{\text{SINR}}\right)\right],
\label{pe}
\end{equation}
where ${\rm Q}(\cdot)$ is the Gaussian $Q$-function \cite[ Eq. (2.1.97)]{pro} and $\beta_M$, $a_p$, and $\tau_M$
are modulation-dependent parameters specified in  \cite{SEP},~\cite{pro}.
%
All parameters in the BEP expression in (\ref{pe}) are deterministic, and the expression is derived based on the Gaussian
distribution of the noise and interference.
However, in the context of cellular networks, many research works have shown that  the aggregate
interference does not follow a Gaussian distribution \cite{renzo2},\!\cite{I2},\!\cite{I3}, thereby rendering (\ref{pe}) obsolete.

One elegant solution  for the exploitation of (\ref{pe}) in the error
performance analysis in  cellular networks is
to assume that each transmitter randomly selects its transmitted
symbol from a Gaussian constellation with unit variance, known as Gaussian signalling approximation, with the main idea of abstracting unnecessary system details (i.e., the interferers' transmitted
symbols) \cite{leila},\!\cite{sawy}. Besides its simplicity, this method  accurately captures the symbol and bit error probabilities
without compromising the modeling accuracy (i.e., does not change the distribution
of the aggregate interference \cite{sawy}). Hereafter, by exploiting the Gaussian signaling
approximation,  we provide the
BEP performance of a typical mobile user on the downlink under the considered channel models, namely  shadowed $\kappa$-$\mu$, $\kappa$-$\mu$, $\eta$-$\mu$, and all other related distributions.
\subsection{General Case}
\textit{Theorem 8: } The average BEP of a cellular downlink over shadowed $\kappa$-$\mu$ fading is
\begin{eqnarray}
{\cal B}^{S\kappa\mu}(\lambda, \alpha )&=&\frac{\beta_M \tau_M}{2}-\frac{ \beta_M \Gamma(\mu\!+\!\frac{1}{2})\sqrt{\!\frac{\Omega}{\mu(1+\kappa)}}}{\sqrt{2}\pi\Gamma(\mu)\left(\frac{\mu\kappa}{m}+1\right)^{m}}\sum_{p=1}^{\tau_M}a_p\int_{0}^{\infty}\frac{{\rm \Psi_1}\left(\mu+\frac{1}{2}, m; \frac{3}{2},\mu;\frac{ -a_p^{2} \xi\Omega}{2\mu(1+\kappa)}, \frac{\mu\kappa}{\mu\kappa+m}\right)}{\sqrt{\xi}}\nonumber\\ && {\cal E}_{r} \left[\exp\left(-\xi r^{\alpha}\frac{\sigma^{2}}{P} \!\right){\cal L}_{I}^{S\kappa\mu}\left(\xi r^{\alpha} \right)\right]\!d\xi.
\label{pkmu}
\end{eqnarray}
\textit{Proof:} See Appendix F.

\textit{Theorem 9: } The average BEP of a cellular downlink  over $\kappa$-$\mu$ fading is
\begin{eqnarray}
{\cal B}^{\kappa\mu}(\lambda, \alpha )&=&\frac{\beta_M\tau_M}{2}-\frac{\beta_M\Gamma(\mu+\frac{1}{2})e^{\kappa\mu}\sqrt{\frac{\Omega}{\mu(1+\kappa)}}}{\sqrt{2}\pi\Gamma(\mu)}\sum_{p=1}^{\tau_M}a_p\int_{0}^{\infty}\frac{{\rm \Psi_2}\left(\mu+\frac{1}{2}; \frac{3}{2},\mu;\frac{ -a_p^{2} \Omega \xi}{2\mu (1+\kappa)},\mu\kappa\right)}{\sqrt{\xi}}\nonumber\\ && {\cal E}_{r} \left[\exp\left(-\xi r^{\alpha}\frac{\sigma^{2}}{P} \right)\!{\cal L}_{I}^{\kappa\mu}\left(\xi r^{\alpha} \right)\right]\!d\xi.
\label{prkmu}
\end{eqnarray}
\textit{Proof:} The result follows after recognizing that ${\cal B}^{\kappa\mu}(\lambda, \alpha )=\lim_{m\rightarrow \infty} {\cal B}^{S\kappa\mu}(\lambda, \alpha )$. Then,  recalling (\ref{limit}) yields the desired result after some manipulations. Note  that the same result could be obtained by following similar steps leading to (\ref{prkmu}) with one difference of using the integral representation of ${\rm \Psi}_2$ in \cite[Eq. 40]{humbert}.
Notice when $\mu=1$ and $\kappa=K$ that (\ref{prkmu}) reduces to the BEP expression under Rice fading.

\textit{Theorem 10: } The average BEP of a cellular downlink over $\eta$-$\mu$ fading is
\begin{eqnarray}
{\cal B}^{\eta\mu}(\lambda, \alpha )&=&\frac{\beta_M\tau_M}{2}-\frac{ \beta_M \eta^{\mu}\Gamma(2\mu+\frac{1}{2})\sqrt{\frac{\Omega\eta}{\mu}}}{\sqrt{2}\pi\Gamma(2\mu)}\sum_{p=1}^{\tau_M}a_p\int_{0}^{\infty}\frac{{\rm \Psi_1}\left(2\mu+\frac{1}{2}, \mu; \frac{3}{2},2\mu;\frac{-a_p^{2} \xi \Omega \eta}{ 2 \mu (1+\eta)}, 1-\eta\right)}{\sqrt{\xi}} \nonumber \\ && {\cal E}_{r} \left[\exp\left(-\xi r^{\alpha}\frac{\sigma^{2}}{P} \right){\cal L}_{I}^{\eta\mu}\left(\xi r^{\alpha} \right)\right]d\xi.
\label{peetamu}
\end{eqnarray}
\textit{Proof:} The average BEP over $\eta$-$\mu$ fading is obtained form  (\ref{pkmu}) by setting $\underline{m}=\mu$, $\underline{\mu}=2\mu$,  and $\underline{\kappa}=\frac{1-\eta}{2 \eta}$ in both the desired and interfering fading channels and performing the necessary simplifications.

\textit{Corollary 5: } The average BEP for the downlink cellular communication
links in general Nakagami-$m$ fading is
\begin{equation}
{\cal B}^{m}(\lambda, \alpha )=\frac{\beta_M\tau_M}{2}-\frac{ \beta_M \Gamma(m+\frac{1}{2})\sqrt{\frac{\Omega}{m}}}{\sqrt{2}\pi\Gamma(m)}\sum_{p=1}^{\tau_M}a_p\int_{0}^{\infty}\!\!\frac{{\rm {}_{1}F_{1}}\left(\!m\!+\!\frac{1}{2}, \frac{3}{2};\frac{ -a_p^{2} \xi\Omega}{2 m}\!\right)}{\sqrt{\xi}}{\cal E}_{r}\! \left[\!\exp\left(\!-\xi r^{\alpha}\frac{\sigma^{2}}{P} \!\right){\cal L}_{I}^{m}\left(\xi r^{\alpha}\! \right)\!\right] d\xi.
\label{pem}
\end{equation}
\textit{Proof:} The result follows  from (\ref{pkmu}) by setting $m=\mu$ and using the Humbert ${\rm \Psi}_1$ function reduction formulas given in Appendix D. Alternatively,  plugging (\ref{mgfnak}) into (\ref{pe}) and following the same steps of Appendix F yield the desired result.

 Recently, the authors of \cite{leila} investigated the impact of Gaussian signalling under Nakagami-$m$  and derived the corresponding error rates. Although the
number of integrals in the obtained BEP expression in (\ref{pem})  is not reduced
when compared to \cite{leila}, our approach discards the necessity for integer $m$,  an assumption made in \cite{leila} for the sake of tractability. In
practical scenarios, however,  the $m$ parameter often takes non-integer
values, as argued by \cite{lorenzo}.  Once again, the  significance of this work is  highlighted by its very wide scope.

It is worthwhile to notice that all the BEP expressions in (\ref{pkmu})-(\ref{pem}) are characterized by the Laplace transforms of the aggregate interference ${\cal L}_{I}$ given in Section II, which are the same ones used to characterize the coverage probability and average rate. As far as computational complexity is concerned, the comparison with previous results is not legitimate since this work is the first of its kind  addressing fading distributions other than Rayleigh and integer Nakgami-$m$ fading. Yet, all the obtained BEP expressions, like those obtained in previous works \cite{sawy1},  \cite{renzo2}, \cite{leila}, encompass a two-fold integration of common built-in functions.
A single integral approximation of the BEP  is only possible for the special case of $\alpha = 4$.  Since that has been extensively investigated in the literature, we omit it  for the sake of conciseness.

\subsection{Special Case of  No Noise and High Signal to Interference Ratio (SIR) }
When interference  dominates the noise (i.e., $\sigma^{2}\longrightarrow0$), the average BEP expressions under the different considered fading distributions follow from (\ref{pkmu})-(\ref{pem}) after computing the expectation over the distance $r$ using \cite[Eq. 3.326.2]{grad}. We are not providing these expressions here since similar proof has been already shown in Section III. However, it is interesting to further push the analysis toward closed-form expressions for the BEP  by considering the high SIR scenario. In this case, to simplify the analysis  we assume that interference undergoes Nakagami-$m$ fading and thus ${\cal L}_{I}$ is given by (\ref{lm}).

Assuming that the desired link undergoes shadowed $\kappa$-$\mu$ fading, then substituting (\ref{lm}) into (\ref{pkmu}) with $\sigma^{2}=0$ and averaging over $r$  yields the interference-limited BEP expression for shadowed $\kappa$-$\mu$ fading on the desired link and Nakagami-$m$ fading on the interfering links as
\begin{equation}
{\cal B}^{S\kappa\mu}( \alpha )=\frac{\beta_M\tau_M}{2}-\frac{ \beta_M \Gamma(\mu+\frac{1}{2})\sqrt{\frac{\Omega}{\mu(1+\kappa)}}}{\sqrt{2}\pi\Gamma(\mu)\left(\frac{\mu\kappa}{m}+1\right)^{m}}\sum_{p=1}^{\tau_M}a_p\int_{0}^{\infty}\frac{{\rm \Psi_1}\left(\mu+\frac{1}{2}, m; \frac{3}{2},\mu;\frac{ -a_p^{2} \xi\Omega}{2\mu(1+\kappa)}, \frac{\mu\kappa}{\mu\kappa+m}\right)}{\sqrt{\xi}{\rm {}_{2}F_{1}}\left(\frac{-2}{\alpha}, m_I, 1-\frac{2}{\alpha}; -\frac{\Omega_I}{m_I}\xi\right)}d\xi.
\label{hkappamu}
\end{equation}

In what follows, we define the average received SIR as SIR$=\frac{\Omega}{\Omega_I}$  and derive  closed-from BEP expressions under the considered fading models when SIR$\rightarrow\infty$.

\textit{Corollary 1 (shadowed $\kappa$-$\mu$, high SIR):} The high SIR interference-limited average BEP  over shadowed $\kappa$-$\mu$ on the desired link and Nakagami-$m$ fading on the interfering links is given by
\begin{eqnarray}
{\cal B}^{S\kappa\mu, \infty}( \alpha )&=&\frac{ \beta_M }{2\left(\frac{\mu\kappa}{m}+1\right)^{m}}\sum_{p=1}^{\tau_M}{\rm \Psi_1}\left(\mu, m; \frac{1}{2},\mu;\frac{a_p^{2}\delta }{4\mu(1+\kappa)}, \frac{\mu\kappa}{\mu\kappa+m}\right)-\frac{ \beta_M \Gamma(\mu+\frac{1}{2})\sqrt{\frac{\delta}{\mu(1+\kappa)}}}{2\Gamma(\mu)\left(\frac{\mu\kappa}{m}+1\right)^{m}}\nonumber\\&& \sum_{p=1}^{\tau_M}a_p{\rm \Psi_1}\left(\mu+\frac{1}{2}, m; \frac{3}{2},\mu;\frac{a_p^{2}\delta }{4\mu(1+\kappa)}, \frac{\mu\kappa}{\mu\kappa+m}\right)
 \label{pskuinf}
\end{eqnarray}
 \textit{Proof:} See Appendix G with $\delta=\frac{(\alpha-2)\Omega}{\Omega_I}$.

\textit{Corollary 2 ($\kappa$-$\mu$, high SIR):}
If the  desired signal fading is $\kappa$-$\mu$ distributed, the average BEP in the  high SIR regime becomes
\begin{eqnarray}
{\cal B}^{\kappa\mu, \infty}( \alpha )&=&\frac{ \beta_M e^{\kappa\mu}}{2}\sum_{p=1}^{\tau_M}{\rm \Psi_2}\left(\mu;  \frac{1}{2}; \mu\frac{a_p^{2}\delta }{4\mu(1+\kappa)}, \mu\kappa\right)-\frac{ \beta_M e^{\kappa\mu} \Gamma(\mu+\frac{1}{2})\sqrt{\frac{\delta}{\mu(1+\kappa)}}}{2\Gamma(\mu)}\nonumber \\&&\sum_{p=1}^{\tau_M}a_p{\rm \Psi_2}\left(\mu+\frac{1}{2}; \frac{3}{2},\mu;\frac{ a_p^{2} \delta}{4 \mu(1+\kappa)},\mu\kappa\right).
\label{hbepkmu}
\end{eqnarray}
\textit{Proof:} The proof starts from (\ref{pskuinf}) and uses the limit relations  in (\ref{limit}) and (\ref{exp}).

\textit{Corollary 3 (Nakagami-$m$, high SIR):}
The high SIR interference-limited average BEP when both the desired and interference channels undergo  Nakagami-$m$ fading reduces to
\begin{eqnarray}
{\cal B}^{m, \infty}( \alpha )&=&\frac{\beta_M}{2}\sum_{p=1}^{\tau_M}\!\!{\rm {}_{1}F_{1}}\!\left(\!m, \frac{1}{2};\frac{ a_p^{2}\delta}{4m }\right)-\frac{ \beta_M \Gamma(m\!+\!\frac{1}{2})\sqrt{\frac{\delta}{ m}}}{2\Gamma(m)}\sum_{p=1}^{\tau_M}\!a_p{\rm {}_{1}F_{1}}\!\left(\!m\!+\!\frac{1}{2}, \frac{3}{2};\frac{ a_p^{2}\delta}{4 m }\right)\ \nonumber\\ &=&
\frac{\beta_M\Gamma(m+\frac{1}{2})}{2 \sqrt{\pi}}\sum_{p=1}^{\tau_M}{\rm U}\left(m, \frac{1}{2}, \frac{ a_p^{2}\delta}{4 m }\right),
\label{hpebm}
\end{eqnarray}
where  ${\rm U}(a,b;z)$ stands for the Tricomi confluent hypergeometric function \cite[ Eq. (9.211.1)]{grad}.

\textit{Proof:} See Appendix H.

Notice that (\ref{hpebm}) also follows  from (\ref{pskuinf}) by setting $m=\mu$ and using the Humbert ${\rm \Psi}_1$ function reduction formulas given in Appendix D,  thereby corroborating again  the correctness of our derivations.

\textit{Remark:} Inspecting (\ref{hpebm}) reveals that the error probability
tends to decrease as $\delta$ and/or  $m$ increase,
meanwhile it is independent of $m_I$. Along the same lines, both ${\cal B}^{S\kappa\mu, \infty}$ and  ${\cal B}^{\kappa\mu, \infty}$ are independent of $m_I$. However,  trends against $\kappa$ and $\mu$ which are rather intricate to investigate analytically will be assessed through simulations in the next section.

\section{NUMERICAL AND SIMULATION RESULTS}
In this section, numerical examples are shown to substantiate
the accuracy of the new unified mathematical framework and to confirm
its potential for analyzing cellular networks. 
All the results shown here have been analytically
obtained by the direct evaluation of the expressions developed
in this paper. Additionally,  using the procedure described in \cite[Sec. V-a]{renzo1}, Monte Carlo simulations have been
performed to validate the derived expressions, and are also
presented in some figures, showing an excellent agreement with
the analytical results. 

 Fig.~1 compares the average rate and average BEP  for the $\kappa$-$\mu$ shadowed fading across a wide range of channel
parameters ($m$, $\kappa$, $\mu$). In Fig.~1~(a), $C^{S\kappa\mu}$ is represented
as a function of the LOS component  in the received wave clusters $\kappa$  for different
values of the $\mu$ parameter. We note that a  rich scattering (large $\mu$)  achieves a higher rate with diminishing returns as
$\kappa$ increases, since  increasing $\mu$  in the strong LOS scenario has little
effect as the performance is dominated by the LOS component. When $m=\mu$, the $\kappa$-$\mu$ shadowed fading distribution boils down to the Nakagami-$m$
distribution, whence the average rate's independency of $\kappa$. 

The impact of shadowed LOS components on performance
can be observed in Fig.~1~(b), where the average rate under $\kappa$-$\mu$ shadowed fading  is presented  as a function of the average SNR
for different values of $m$  and considering, respectively, small ($\kappa=1$) and large ($\kappa=20$)
LOS components. It can be
observed that performance improves with
small LOS components ($\kappa= 1$) if the latter are affected by
heavy shadowing (small $m$). However, when the shadowing is mild, large
LOS components ($\kappa= 20$) always  improve the average rate.
In fact,  small $m$ indicates highly fluctuating dominant components,
which decrease both the received signal and the aggregate interference powers thereby
increasing the SINR level and ultimately achieving higher rates. Conversely, when $m$ is large, the shadowing on the dominant components subsides and $\kappa$-$\mu$ shadowed fading
reduces to $\kappa$-$\mu$ fading. Moreover, light shadowing always yields higher interference power  thereby deteriorating the received SINR
level as well as the average rate. Fig.~1~(b) also compares the average rate for various BS densities $\lambda$. It can be seen that the average
rate of a sparse network  ($\lambda \leq 10^{-4}$) is much lower than that of a dense network ($\lambda \geq 10^{-2}$). For example the
average average rate is about $0.02$ for $\lambda = 10^{-4}$ and $1$ for $\lambda= 10^{-2}$ with $m=0.5$, $\kappa=20$  and $\text{SNR} = 15$ dB.

Fig.~1~(c) plots the average BEP of the downlink with coherent QPSK ($M=4$) under shadowed $\kappa$-$\mu$  fading on both the desired and interfering links.
 As can be seen, a strong dominant
LOS component (large $\kappa$) and rich scattering (large $\mu$) collectively improve the error performance. The figure also shows that Rician fading on the useful
link has higher performance than the Rayleigh fading due to
the LOS path.

Fig.~1~(d) shows the average BEP versus the received SIR
for various shadowed $\kappa$-$\mu$ fading environments in downlink. Recall that
although the Rician model represents a LOS (and better)
channel than the Rayleigh model, the large SNR tails of the
Rician distribution always have the same decay rate as (i.e., are
parallel to) the tails of the Rayleigh distribution (reflecting the
diversity order of the channel). At high SIR,  the asymptotic expansion in (\ref{pskuinf}) matches very well its exact counterpart, which confirms the validity of our mathematical analysis for different parameter settings.

\begin{figure}
\centering
\hspace{-0.2cm}
\subfigure[ $C^{S\kappa\mu}$ versus  $\kappa$ for  $m=m_I=3$, and $\mu_I=1$. ]{
            \includegraphics[width=7.6cm] {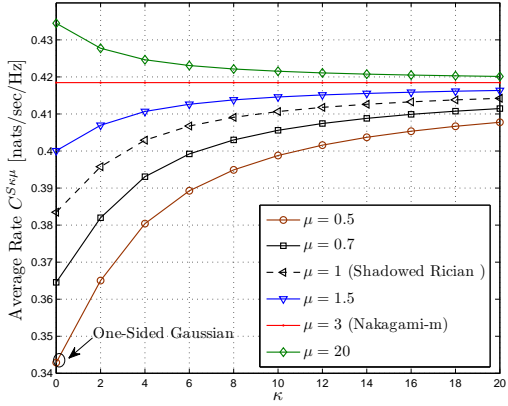}

}
\hspace{-0.2cm}
\subfigure[${\cal C}^{S\kappa\mu}$ versus SNR for different values of $\kappa=\kappa_I$ and $m=m_I$ with $\mu=\mu_I=2$. ]{
            \includegraphics[width=7.6cm] {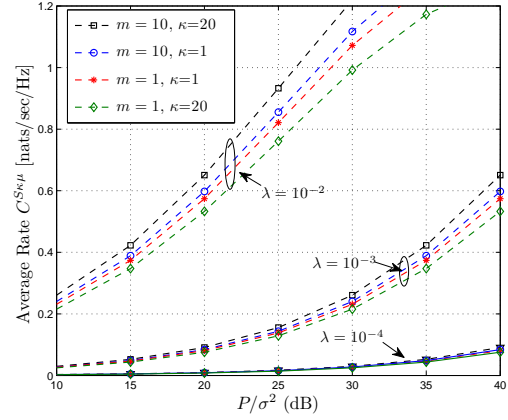}
}
\hspace{-0.2cm}
\subfigure[ ${\cal B}^{S\kappa\mu}$  versus  $\kappa$ for  $m=m_I=3$,  $\mu_I=1$,  and $M = 4$.]{
            \includegraphics[width=7.6cm] {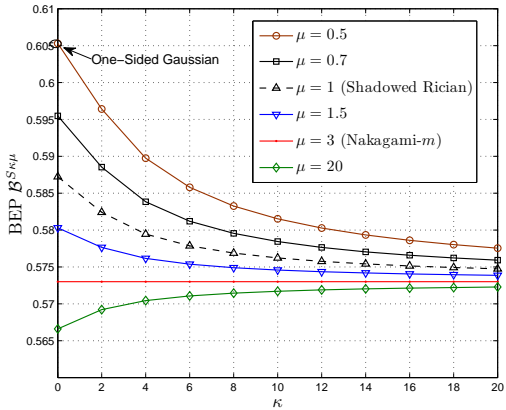}

}
%
   \hspace{-0.2cm}
\subfigure[${\cal B}^{S\kappa\mu}$ and ${\cal B}^{S\kappa\mu, \infty}$ with Rayleigh interference versus SIR for $M\in \{4,16\}$. ]{
            \includegraphics[width=7.6cm] {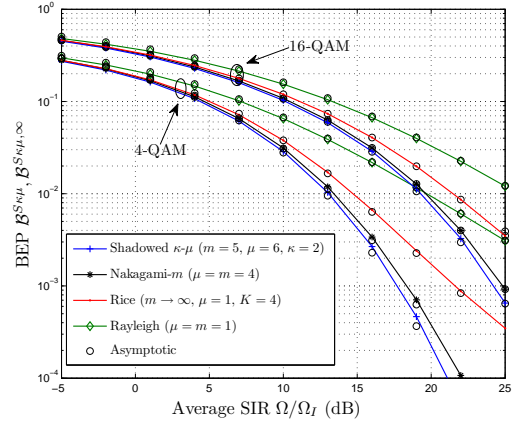}
   }
\hspace{-0.2cm}
\caption{Performance of downlink transmission over shadowed $\kappa$-$\mu$ fading. Setup: $\Omega_I=\Omega$, $\kappa_I=\kappa$, $\lambda = 10^{-4}$,  and $\alpha=3$.}
\label{app2sim}
\end{figure}

\begin{figure}
\centering
\hspace{-0.2cm}
\subfigure[ $C^{\kappa\mu}$  versus $\kappa$ with $\mu_I=1$.]{
            \includegraphics[width=7.5cm] {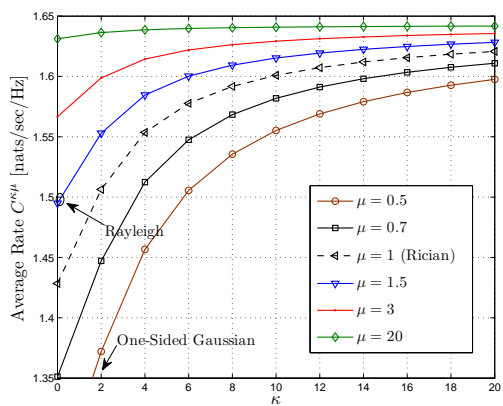}

}
\hspace{-0.2cm}
\subfigure[$C^{\kappa\mu}$  versus BS density $\lambda$ for different values of $\mu=\mu_I$ with $\kappa=1.5$. ]{
            \includegraphics[width=7.5cm] {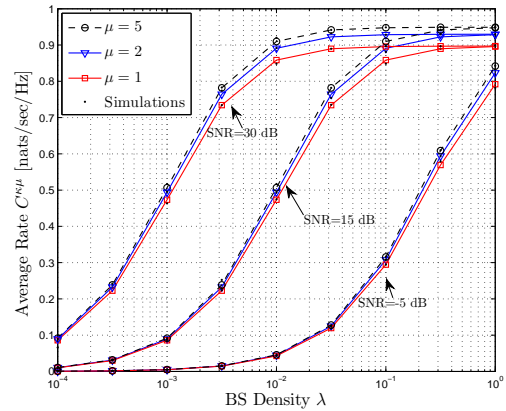}

}
\hspace{-0.2cm}
\subfigure[SIR coverage probability $P_{cov}^{\kappa\mu}$.]{
            \includegraphics[width=7.5cm] {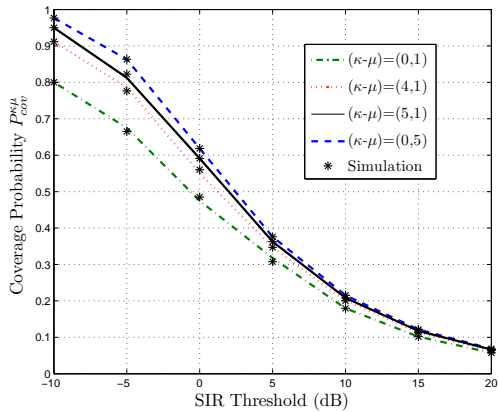}

}
\hspace{-0.2cm}
\subfigure[${\cal B}^{\eta\mu}$ versus $\eta$ for $M = 4$ with $\mu_I=1$.]{
            \includegraphics[width=7.5cm] {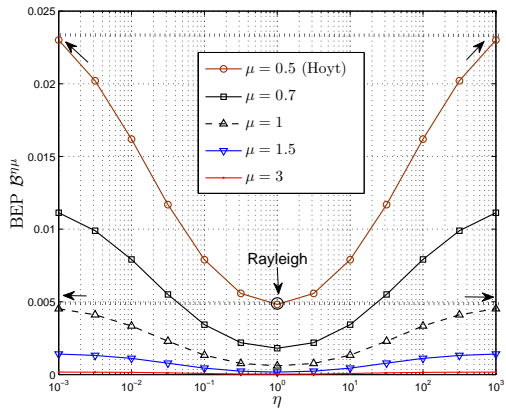}
   }
    
   \caption{Performance of downlink transmission over $\kappa$-$\mu$  and $\eta$-$\mu$ fading. Setup: $\Omega_I=\Omega$, $\kappa_I=\kappa$, $\lambda = 10^{-4}$, and $\alpha=3$.}
\label{app2sim}
\end{figure}

In Figs.~2~(a)-(d), the performance of downlink transmission over $\kappa$-$\mu$  and $\eta$-$\mu$ fading is presented.\\Fig.~2~(a) depicts the ergodic rate $C^{\kappa\mu}$  under the $\kappa$-$\mu$ fading model. Since  the shadowing can be neglected in this case, any increase in the power of the dominant
components is obviously favorable for the channel capacity. It is therefore straightforward to see that increasing the
parameter $\kappa$ implies increasing the ergodic rate since a higher LOS power implies improving the capacity of
the $\kappa$-$\mu$ channel.

Fig.~2~(b)   compares the average rate under $\kappa$-$\mu$ fading   versus the BS density $\lambda$  for different
values of the  $\mu$  parameter. As conjectured in Corollay 1,  the network performance is invariant of the
network density in an  interference-limited
condition (large BS intensity). The results show that
 the rate saturation may happen at certain network density required to obtain sufficiently larger interference power than the noise. In fact, at high SNR,   the saturation regime is reached at $\lambda = 10^{-2}$, compared to $\lambda\geq 10^{-1}$ in the low SNR regime.
In practice, installing more BSs is beneficial to the user
performance up to a density point, after which further
densification turns out to be extremely ineffective  due
to faster growth of interference compared to useful signal.
This highlights the cardinal importance of interference
mitigation, coordination among neighboring cells and
local spatial scheduling.

Fig.~2~(c) plots the coverage probability versus the SIR threshold
for  different fading environments contained within the $\kappa$-$\mu$ model.  It can be seen  that both $\mu$ and $\kappa$ have impact on the coverage probability with a more pronounced impact for the cluster number $\mu$, especially for small $\kappa$, a behavior previously observed in Fig.~1~(a).

Fig.~2~(d) depicts the interference-limited average BEP for $\eta$-$\mu$ fading, which is, as expected, symmetrical on a logarithmic scale around the minimum value of the average BEP at $\eta=1$ regardless of the number of
clusters $\mu$. In the same figure, we have specified the limit cases
for $\eta\rightarrow0$ and $\eta\rightarrow \infty$. When $\mu = 0.5$, the $\eta$-$\mu$ model collapses into
the one-sided Gaussian model for $\eta = 0$ or $\eta\rightarrow \infty$, whereas
for $\eta=1$, it collapses into the Rayleigh model. When $\mu = 1$,
the $\eta$-$\mu$ is reduced to the Rayleigh case for $\eta = 0$ or $\eta\rightarrow \infty$.
Both the Rayleigh and one-sided Gaussian BEP  values are illustrated by  horizontal dotted
and dashed lines, respectively.

In Fig.~1 and Fig.~2, please note that we have identified in the figure legends or with rounded circles the performance curves or points, respectively, of some particular fading distributions that simply reduce from the shadowed  $\kappa$-$\mu$, the $\kappa$-$\mu$, and the $\eta$-$\mu$ models.

\section{CONCLUSION}
In this paper a novel mathematical methodology for performance
evaluation on the downlink of cellular networks over fading channels is presented. The proposed approach exploits results from stochastic geometry for the computation of the SINR's MGF,
 which is shown to be conveniently formulated as a function of a desired-user fading dependent term and the Laplace transform of the interference.
By capitalizing on
this mathematical formulation, we have been able to obtain  two-fold integral expressions for
the ergodic rate, the coverage probability, and the tangible decoding error probability  for various fading distributions.
Remarkably, the proposed framework accommodates
generic fading distribution models including shadowed $\kappa$-$\mu$, $\kappa$-$\mu$, and $\eta$-$\mu$ that account for LOS/NLOS and shadowed fading.
Our results provide useful insights into the coverage, throughput, and BEP performance in complex fading environments and shed new lights on  the prominent impact of  DSCs and shadowed DSCs
propagation on cellular networks. Finally, this new framework is flexible to capture several fading
conditions ranging from deterministic and favorable Rician
to severely shadowed and Rayleigh fading. Future work possibilities relying on this new modeling paradigm are tremendous and include without limitation the extension
to  multi-tier heterogeneous  networks (HetNet)s with arbitrary numbers of tiers having different densities and  transmit powers. Moreover, the developed
baseline model  provides new  powerful tools to analyze other
network architectures such that device-to-device (D2D) and MIMO-enabled networks.

\section{Appendix}
\subsection{Proof of Lemma 1:}

Given the $\text{SINR}=\frac{  h r^{-\alpha}}{\frac{\sigma^{2}}{P} +{\cal I}(r)}$, its MGF, defined as  $M_{\text{SINR}}(s)\triangleq{\cal E}_{r,h,{\cal I} }\left[\exp\left(-s \text{SINR}\right)\right]$, is evaluated as
\begin{eqnarray}
M_{\text{SINR}}(s)&=& {\cal E}_{r,h}\left[\int_0^{\infty} \exp\left(-\frac{ s h r^{-\alpha}}{y}\right)f_{{\cal I}+\frac{\sigma^{2}}{P}}(y)dy\right]\nonumber\\
&=&{\cal E}_{r,h}\left[{\cal L}_{{\frac{1}{I+\frac{\sigma^{2}}{P}}}}\left(s h r^{-\alpha}\right)\right]\nonumber\\
&\overset{(a)}{=}&{\cal E}_{r}\left[1-2\sqrt{s  r^{-\alpha}}\int_{0}^{\infty}{\cal E}_{h}\left[\sqrt{h}{\rm J}_1\left(2\sqrt{s h  r^{-\alpha}} \xi\right)\right]{\cal L}_{{\cal I}+\frac{\sigma^{2}}{P}}(\xi^{2}) d\xi\right]
\end{eqnarray}
where  $(a)$ follows from applying \cite[Theoerm 1]{theo} and ${\rm J}_1(\cdot)$ is the Bessel function of the first kind and first order \cite[Eq. 8.402]{grad}.

\subsection{Proof of $M^{S\kappa\mu}_{\text{SINR}}$ and ${\cal L}_{I}^{S\kappa\mu}$:}

After applying ({\ref{mgf}), the expectation over shadowed $\kappa$-$\mu$ fading using the distribution in (\ref{pdfsku})  can be calculated as
\begin{equation}
{\cal E}_{h}\left[\sqrt{h}{\rm J}_1\left(2\sqrt{s h  r^{-\alpha}} \xi\right)\right]= A \int_{0}^{\infty} y^{\mu-\frac{1}{2}}e^{-\emph{B}y}{\rm J}_1\left(2\sqrt{s y  r^{-\alpha}} \xi\right){\rm {}_{1}F_{1}}\left(m, \mu , \emph{C} y \right)dy,
\end{equation}
where we denote
$A= \frac{ \mu^{\mu} m^{m}(1+\kappa)^{\mu}}{\Gamma(\mu)\Omega^{\mu} (\mu \kappa+m)^{m}}$, $B= \frac{ \mu(1+\kappa)}{\Omega}$, and $C= \frac{ \mu^{2} \kappa(1+\kappa)}{\Omega(\mu \kappa+m)}$.
Recalling the Bessel ${\rm J}_{\nu}(\cdot)$ representation through the more general confluent hypergeometric function ${\rm {}_{1}F_{1}}(\cdot)$ given by
\begin{equation}
{\rm J}_{\nu}(z)=\frac{z^{\nu}}{2^{\nu}\Gamma(\nu+1)}\lim_{a\rightarrow\infty}{\rm {}_{1}F_{1}}\left(a, \nu+1; \frac{-z^{2}}{4 a}\right),
\end{equation}
 then it follows that
\begin{eqnarray}
{\cal E}_{h}\left[\sqrt{h}{\rm J}_1\left(2\sqrt{s h  r^{-\alpha}} \xi\right)\right]
&=&A \sqrt{s  r^{-\alpha}} \xi \lim _{a\rightarrow\infty} \int_{0}^{\infty} y^{\mu}e^{-\emph{B}y}{\rm {}_{1}F_{1}}\left(a, 2 ,  \frac{- s  r^{-\alpha} \xi^{2}}{a} y \right){\rm {}_{1}F_{1}}\left(m, \mu , \emph{C} y \right)dy \nonumber \\
 &\overset{(a)}{=}&A \sqrt{s r^{-\alpha}} \xi B^{-\mu-1}\Gamma(\mu+1) \lim _{a\rightarrow\infty}{\rm F_{2}}\left(\mu+1,a,m,2,\mu;\frac{- s  r^{-\alpha} \xi^{2}}{a B}, \frac{C}{B}\right)\nonumber \\
&\overset{(b)}{=}& \frac{A\Gamma(\mu+1)}{B^{\mu+1}} \sqrt{s r_0^{-\alpha}} \xi  {\rm \Psi_1}\left(\mu+1, m; 2,\mu;\frac{- s  r^{-\alpha} \xi^{2}}{ B}, \frac{C}{B}\right),
\label{eh}
\end{eqnarray}
where $(a)$ follows from recognizing Appell's  ${\rm F}_{2}$ representation \cite[Eq. 27]{humbert}
\begin{equation}
{\rm F_{2}}\left(a,b,b',c,c';\frac{w}{p},\frac{ z}{p}\right)=\frac{p^{a}}{\Gamma(a)}\int_{0}^{\infty}x^{a-1}e^{- p x}{\rm {}_{1}F_{1}}\left(b, c ,   w x \right){\rm {}_{1}F_{1}}\left(b', c' ,    x z \right)dx, \quad {\rm Re}(a), {\rm Re}>0,
\end{equation}
and $(b)$ is obtained by limit (confluence) formula \cite{humbert}
\begin{equation}
\lim_{\epsilon\rightarrow0} {\rm F_{2}}\left(\alpha,\frac{b'}{\epsilon},b,c',c;\epsilon x, y\right)={\rm \Psi_1}\left(\alpha, b; c',c; b'x , y\right).
\label{limit1}
\end{equation}
Finally substituting (\ref{eh}) into (\ref{mgf}) and carrying out the change of $\xi =r^{-\alpha} \xi^{2}$ yield the desired result after some manipulations,  notably taking into account the linearity and the time shifting
properties  of the Laplace transform implying that ${\cal L}_{{\cal I}+\sigma^{2}}(x)=e^{- \sigma^{2} x}{\cal L}_{{\cal I}}(x)$.

The Laplace transform of the aggregate
interference from the interfering BSs received at the tagged user under $\kappa$-$\mu$ shadowed fading, denoted as  ${\cal L}_{I}^{S\kappa\mu}(\xi )$,
is obtained as
\begin{eqnarray}
{\cal L}_{{\cal I}}(\xi)&=& {\cal E}_{\Psi, h}\left[\exp\left(-\xi \sum_{i\in\Psi^{(\setminus 0)}} h_i r_i^{-\alpha}\right)\right]\nonumber\\
&\overset{(a)}{=}& \exp\left(-2 \pi \lambda \int_{r}^{\infty}\left(1-{\cal E}_{h}[\exp\left(-\xi  h v^{-\alpha}\right)]\right)v dv\right)\nonumber\\
&\overset{(b)}{=}& \exp\left(-2 \pi \lambda \int_{r}^{\infty}\left(1-\frac{\mu_I^{m_I} m_I^{m_I}(1+
\kappa_I)^{\mu_I}}{\Omega_I^{\mu_I} (\mu_I \kappa_I+m_I)^{m_I}}\frac{\left(\xi v^{-\alpha}+\frac{\mu_I(1+\kappa_I)}{\Omega_I}\right)^{m_I-\mu_I}}{\left(\xi v^{-\alpha}+\frac{\mu_I(1+\kappa_I)}{\Omega_I}\frac{m_I}{\mu_I \kappa_I+m_I}\right)^{m_I}}\right)v dv\right),
\label{agg}
\end{eqnarray}
where $(a)$ follows from independence between $\Psi$ and $h_i$
and  the probability generation functional
(PGFL) of the PPP \cite{bacelli} and  $(b)$ follows from  the MGF of $h_i$ under shadowed $\kappa$-$\mu$ fading  recently derived in \cite[Eq. 5]{paris}.
Further by letting $x\longleftarrow r^{\alpha} v^{-\alpha}$ in (\ref{agg}), the latter is obtained as
\begin{equation}
{\cal L}_{I}^{S\kappa\mu}(\xi r^{\alpha})=\exp\Bigg(-2 \pi \lambda \frac{r^{2}}{\alpha} \int_{0}^{1}x^{-\frac{2}{\alpha}-1}\left(1-\frac{\left(1+\frac{\mu_I(1+\kappa_I)}{\Omega_I} \xi x\right)^{m_I-\mu_I}}{\left(1+\frac{ (\mu_I\kappa+m_I)\Omega_I}{m_I\mu_I(1+\kappa_I)} \xi x\right)^{m_I}}\right)dx\Bigg),
\label{m1}
\end{equation}
 when $\mu_I\leq m_I$, and in  case of  $\mu_I\geq m_I$, it can be calculated as
\begin{equation}
{\cal L}_{I}^{S\kappa\mu}(\xi r^{\alpha})=\exp\Bigg(-2 \pi \lambda \frac{r^{2}}{\alpha} \int_{0}^{1}x^{-\frac{2}{\alpha}-1}\left(1-\frac{1}{\left(1+\frac{\mu_I(1+\kappa_I)}{\Omega_I} \xi x\right)^{\mu_I-m_I}\left(1+\frac{ (\mu_I\kappa+m_I)\Omega_I}{m_I\mu_I(1+\kappa_I)} \xi x\right)^{m_I}}\right)dx\Bigg).
\label{m2}
\end{equation}
Let $\Theta=\frac{\mu_I(1+\kappa_I)}{\Omega_I}$ and $\Xi=\frac{ (\mu_I\kappa_I+m_I)\Omega_I}{m_I\mu_I(1+\kappa_I)}$. Then applying binomial expansion on $\left(\!1\!+\!\Xi\xi x\!\right)^{m_I}\!-\!\left(\!1\!+\!\Theta \xi x\!\right)^{m_I-\mu_I}\!\!=\!\!\sum_{k=1}^{m_I}\binom{m_I}{k}\Xi^{k}\xi^{k}-\sum_{n=1}^{m_I-\mu_I}\binom{m_I-\mu_I}{n}\Theta^{n}\xi^{n}$ in the numerator  of (\ref{m1}) and on  $\left(1\!+\!\Xi\xi x\right)^{m_I}\!\!\left(1\!+\!\Theta \xi x\right)^{m_I-\mu_I}\!\!=\!\!\sum_{k=0}^{m_I} \sum_{n=0}^{m_I-\mu_I}\binom{m_I}{k}\binom{m_I-\mu_I}{n}\Xi^{k}\Theta^{n}x^{k+n}$ in the numerator
of (\ref{m2}), we obtain \footnote{Note that the obtainment of (\ref{m3}) and (\ref{m4})  inflicts the quantities $m_I$ and $\mu_I$ to be integer valued.}
\begin{eqnarray}
{\cal L}_{I}^{S\kappa\mu}(\xi r^{\alpha})= \exp\Bigg(-2 \pi \lambda \frac{r_0^{2}}{\alpha}\Bigg( \sum_{k=1}^{m_I}\binom{m_I}{k}\Xi^{k}\xi^{k}\int_{0}^{1}\frac{x^{k-\frac{2}{\alpha}-1}}{(1+\Xi \xi x)^{m}}dx\nonumber \\ -
\sum_{n=1}^{m_I-\mu_I}\binom{m_I-\mu_I}{n}\Theta^{n}\xi^{n}\int_{0}^{1}\frac{x^{n-\frac{2}{\alpha}-1}}{(1+\Xi \xi x )^{m}}dx\Bigg)\Bigg) , \quad\text{when }\mu_I\leq m_I,
\label{m3}
\end{eqnarray}
and
\begin{eqnarray}
{\cal L}_{I}^{S\kappa\mu}(\xi r^{\alpha})=\exp\Bigg(-2 \pi \lambda \frac{r^{2}}{\alpha} \sum_{n,k; (n,k)\neq(0,0)}^{m_I, \mu_I-m_I }\binom{\mu_I-m_I}{k}\binom{m_I}{n}\Theta^{k} \Xi^{n}\xi ^{k+n}\nonumber \\ \int_{0}^{1}\frac{x^{k+n-\frac{2}{\alpha}-1}}{\left(1+\Theta \xi x\right)^{\mu_I-m_I}\left(1+\Xi \xi x\right)^{m_I}}dx\Bigg), \quad \text{when }\mu_I\geq m_I.
\label{m4}
\end{eqnarray}
Closed-form expressions of (\ref{m3}) and (\ref{m4}) are obtained after recognizing that
\begin{equation}
{\rm {}_{2}F_{1}}\left(a, b;c; x\right)=\frac{\Gamma(c)}{\Gamma(b)\Gamma(c-b)}\int_{0}^{1}t^{b-1}(1-t)^{c-b-1}(1- t x)^{-a}dt, \quad {\rm Re}(c)> {\rm Re}(b)>0,
\label{F2}
\end{equation}
and
\begin{equation}
{\rm F_{1}}\left(a, b, b'; c; w,z\right)=\frac{\Gamma(a)}{\Gamma(c)\Gamma(c-a)}\int_{0}^{1}t^{a-1}(1-t)^{c-a-1}(1- t w)^{-b}(1- t z)^{-b'}dt, \quad {\rm Re}(a)>0,
\end{equation}
which completes the proof and provide explicitly the ${\cal L}_{I}^{S\kappa\mu}(\xi )$  as shown in \textit{Theorem 1}.

\subsection{Proof of $M^{\kappa\mu}_{\text{SINR}}$ and ${\cal L}_{I}^{\kappa\mu}$: }
The $\kappa$-$\mu$ fading arises from the shadowed $\kappa$-$\mu$ fading as $m\rightarrow\infty$. Accordingly, it follows that $
M^{\kappa\mu}_{\text{SINR}}=\lim_{m\rightarrow\infty}M^{S\kappa\mu}_{\text{SINR}}$,  and ${\cal L}_{I}^{\kappa\mu}=\lim_{m_I\rightarrow\infty} {\cal L}_{I}^{S\kappa\mu}$.
As far as $M^{\kappa\mu}_{\text{SINR}}$  is concerned, the desired result follows by applying  the following properties:
\begin{equation}
\lim_{\epsilon\rightarrow0}{\rm \Psi_1}\left(a, \frac{b}{\epsilon}; c, c'; \epsilon w , z\right)={\rm \Psi_2}\left(a; c, c'; b w , z\right),
\label{limit}
\end{equation}
and
\begin{equation}
\lim_{a\rightarrow\infty}\left(\frac{x}{a}+1\right)^{-a}= e^{-x},
\label{exp}
\end{equation}
where (\ref{exp}) is the well-known limit that defines the
exponential function.

Regarding ${\cal L}_{I}^{\kappa\mu}$, the specialization from ${\cal L}_{I}^{S\kappa\mu}$ is not straightforward and requires further manipulations.
The proof tracks the proof of ${\cal L}_{I}^{S\kappa\mu}$  up until step $(a)$
of  (\ref{agg}). Then,
\begin{eqnarray}
{\cal L}_{\cal I}^{\kappa\mu}(\xi)&\overset{(a)}{=}& \exp\left(-2 \pi \lambda \int_{r}^{\infty}{\cal E}_{h}\left[\xi  h v^{-\alpha}\exp\left(-\xi  h v^{-\alpha}\right){\rm {}_{1}F_{1}}\left(1, 2; \xi  h v^{-\alpha}\right)\right]v dv\right)\nonumber\\
&\overset{(b)}{=}& \exp\Bigg(-2 \pi \lambda \xi \frac{ \mu_I (1+\kappa_I)^{\frac{\mu_I+1}{2}}}{e^{\kappa_I\mu_I}\Omega_I^{\frac{\mu_I+1}{2}} \kappa_I^{\frac{\mu_I-1}{2}}} \int_{r}^{\infty}\frac{v^{-\alpha+1}}{\left(\xi v^{-\alpha}+\frac{\mu_I(\kappa_I+1)}{\Omega_I}\right)^{\frac{\mu_I-1}{2}}}\nonumber \\ && \!\!\!\!\!\!\!\!\!\!\!\!\int_{0}^{\infty}\!\!\! x^{\frac{\mu_I+1}{2}} e^{-x}\! {\rm {}_{1}F_{1}}\!\left(\!1, 2; \frac{\xi   v^{-\alpha}}{\xi v^{-\alpha}+\frac{\mu_I(\kappa_I+1)}{\Omega_I}}x\!\right)\!{\rm I}_{\mu_I-1}\left(\!2 \mu_I\sqrt{\frac{\kappa_I(1+\kappa_I)x}{\Omega_I \left(\xi v^{-\alpha}+\frac{\mu_I(\kappa_I+1)}{\Omega_I}\!\right)}}\!\right) \!dx dv\!\Bigg),
\end{eqnarray}
where $(a)$ follows from using the relation $(1-e^{-x})/x=  e^{-x}{\rm {}_{1}F_{1}}\left(1, 2; x\right)$ and $(b)$ follows from the $\kappa$-$\mu$ distribution of $h_i$ given in (\ref{pdfku}) and carrying out the change of variable $x=h\left(\xi v^{-\alpha}+\frac{\mu_I(\kappa_I+1)}{\Omega_I}\right)$.
\begin{eqnarray}
{\cal L}_{\cal I}^{\kappa\mu}(\xi r^{\alpha})\!\!\!\!\!&\overset{(c)}{=}&\!\!\!\!\exp\!\!\Bigg(\!\!\!-2 \pi \lambda  \frac{ \mu_I^{2} (1\!+\!\kappa_I)^{\mu_I}}{e^{\kappa_I\mu_I}\Omega_I^{\mu_I} }\!\! \int_{0}^{1}\frac{\xi v^{-\alpha+1}{\rm \Psi_1}\left(\!\mu_I\!+\!1, 1; \mu_I,2; \frac{\mu_I^{2}\kappa_I(1\!+\!\kappa_I)}{\Omega_I \left(\xi x+\frac{\mu_I(\kappa_I+1)}{\Omega_I}\right)} , \frac{\xi   x}{\xi x+\frac{\mu_I(\kappa_I+1)}{\Omega_I}}\right)}{\left(\xi x+\frac{\mu_I(\kappa_I+1)}{\Omega_I}\right)^{\mu_I-1}}dx\Bigg)\nonumber\\
&\overset{(d)}{=}& \!\!\!\!\exp\!\!\Bigg(\!\!\!-2 \pi \lambda  \frac{r^{2} \mu_I^{2} (1\!+\!\kappa_I)^{\mu_I}}{\alpha e^{\kappa_I\mu_I}\Omega_I{\mu_I} }\!\!\sum_{k=0}^{\infty}\!\!\frac{(\mu_I\!+\!1)_k\left(\frac{\mu_I^{2}\kappa_I(1+\kappa_I)}{\Omega_I}\right)^{k}}{k!(\mu_I)_k (\mu_I+k)}\int_{0}^{1}\frac{ x^{-\frac{2}{\alpha}-1}\Big(\Big(\frac{\xi x}{\frac{\mu_I(\kappa_I+1)}{\Omega_I}}+1\Big)^{\mu_I+k}\!\!\!-\!1\Big)}{\left(\xi x\!\!+\!\frac{\mu_I(\kappa_I+1)}{\Omega_I}\right)^{\mu_I+k-2}}dx\Bigg)\nonumber\\
&\overset{(e)}{=}&\!\!\!\!\exp\!\!\Bigg(\!\!\!-2 \pi \lambda  \frac{r^{2} \mu_I^{2} (1\!+\!\kappa)^{\mu_I}}{\alpha e^{\kappa_I\mu_I}\Omega_I^{\mu_I} }\!\!\sum_{k=0}^{\infty}\frac{(\mu_I+1)_k\left(\frac{\mu_I^{2}\kappa_I(1+\kappa_I)}{\Omega_I}\right)^{k}}{k!(\mu_I)_k (\mu_I+k)}\sum_{n=1}^{\mu_I+k}\frac{\binom{\mu_I+k}{n}\left(\frac{\xi\Omega_I}{\mu_I\kappa_I(1+\kappa_I)}\right)^{n}}{n-\frac{2}{\alpha}}\nonumber \\ && {\rm {}_{2}F_{1}}\left(\mu_I+k-2,n-\frac{2}{\alpha},n+1-\frac{2}{\alpha},-\frac{\xi\Omega_I}{\mu_I\kappa_I(1+\kappa_I)}\right)\Bigg),
\label{aggkmu}
\end{eqnarray}
In (\ref{aggkmu}), $(c)$  results in the same line of (\ref{eh}) after using $
{\rm I}_{\nu}(z)=\frac{z^{\nu}}{2^{\nu}\Gamma(\nu+1)}\lim_{a\rightarrow\infty}{\rm {}_{1}F_{1}}\left(a, \nu+1; \frac{z^{2}}{4 a}\right)$  followed by the change of variable $x= \left(\frac{r}{v}\right)^{\alpha}$. The aggregate interference under $\kappa$-$\mu$ still needs manipulations as to solve the integral involving the Humbert function obtained after $(c)$. To this end, we resort  in $(d)$ to the the series expansion of the Humbert function ${\rm \Psi_1}$ given by \cite{humbert}
\begin{equation}
{\rm \Psi_1}\left(a, b; c,c'; x , y\right)=\sum_{k=0}^{\infty}\frac{(a)_k}{k! (c')_k} y^{k} {\rm {}_{2}F_{1}}\left( a+k, b, c; x\right), \quad |x|<1,
\end{equation}
where $(a)_n$ denote the Pochhammer symbol,  along with the reduction formulas of the Gauss hypergeometric function ${\rm {}_{2}F_{1}}(\cdot)$ given in \cite[Eq. 9.121.5]{grad}. Finally $(e)$ follows from using to the binomial expansion and recognising (\ref{F2}), thereby leading the desired result after some manipulations.
\subsection{Proof of $M^{m}_{\text{SINR}}$ and ${\cal L}_{I}^{m}$: }
When $m=\mu$,  it holds that
\begin{eqnarray}
{\rm \Psi_1}\left(m+1, m; 2,m;\frac{- s \xi \Omega}{ m(1+\kappa)}, \frac{m \kappa}{m \kappa+m}\right)\!\!\!\!&\overset{(a)}{=}&\!\!\!\lim_{\epsilon\rightarrow0} \left(1\!-\!\frac{m \kappa}{m \kappa+m}\right)^{-m-1}\!\!{\rm {}_{2}F_{1}}\left(m+1, \frac{\beta}{\epsilon}; 2; \frac{\frac{- s \epsilon \xi \Omega}{ m(1+\kappa)}}{1-\frac{m \kappa}{m \kappa+m}}\right) \nonumber\\&\overset{(b)}{=}&\left(1-\frac{m \kappa}{m \kappa+m}\right)^{-m-1}{\rm {}_{1}F_{1}}\left(m+1 , 2; \frac{\frac{- s  \xi \Omega}{ m(1+\kappa)}}{1-\frac{m \kappa}{m \kappa+m}}\right),
\label{psim}
\end{eqnarray}
where $(a)$ follows form using (\ref{limit1}) and applying the reduction formulas \cite[Eq. 9.182.2]{grad}
\begin{equation}
{\rm F_{2}}\left(\alpha,\beta,b,c,b;x, y\right)=(1-x)^{-\alpha}{\rm {}_{2}F_{1}}\left(\alpha, \beta; c; \frac{y}{1-x}\right),
\end{equation}
and $(b)$ follows from evaluating the limit according to $\lim_{\epsilon\rightarrow0}{\rm {}_{2}F_{1}}\left(\alpha, \frac{b'}{\epsilon}; c'; \epsilon z\right)={\rm {}_{1}F_{1}}\left( b'; \gamma; z\right)$.
Substituting  (\ref{psim}) into (\ref{MGFSKMU}) yields the desired result after some simplifications.
The Laplace transform of the aggregate interference under Nakagami-$m$ fading, i.e.,   ${\cal L}_{I}^{m}$,   specialises from  ${\cal L}_{I}^{S\kappa\mu}$ when $m_I=\mu_I$. In this case, its is straightforward to show that
the second summation in the RHS of (\ref{L1}) vanishes  while the first summation reduces to ${\cal L}_{I}^{m}$.

\subsection{Proof of $C^{m}$:}
Setting $m=\mu$  in  (\ref{Cnak}) and resorting to \cite[Theorem 1]{appel}, we have
\begin{eqnarray}
{\rm F_2}\left(\!\!\mu\!+\!1, m, 1;\mu,2;\frac{\mu \kappa}{\mu \kappa+m},\frac{-  \xi \Omega}{ \mu(1\!+\!\kappa)}\!\! \right)\!\!\!\!\!&=&\!\!\!\!\frac{(1\!+\!\kappa)}{   \xi \Omega}\!\!{\rm {}_{2}F_{1}}\left(\!\! \mu, m; \mu; \frac{\mu \kappa}{\mu \kappa\!+\!m}\!\!\right)\!-\!\frac{(1\!+\!\kappa){\rm {}_{2}F_{1}}\left(\!\! \mu, m; \mu; \frac{\frac{\mu \kappa}{\mu \kappa+m}}{1+\frac{  \xi \Omega}{ \mu(1+\kappa)}}\right)}{ \xi \Omega (1\!+\!\frac{ \xi \Omega}{ \mu(1+\kappa)})^{\mu}}\nonumber \\ &\overset{(b)}{=}&\frac{(\kappa+1)^{m+1}}{\Omega \xi}\left(1-\left(1+\frac{\xi \Omega}{m}\right)^{-m}\right),
\label{simp}
\end{eqnarray}
 where  $(b)$ follows form applying ${\rm {}_{2}F_{1}}\left( a,b; b; z\right)=(1-z)^{-a}$. Substituting (\ref{simp}) into (\ref{Cnak}) yields the desired result after some manipulations.

\subsection{Proof of ${\cal B}^{S\kappa\mu}(\lambda,\alpha)$}
Using Craig's alternative expression for the Gaussian $Q$-function
\cite[Eq. 9]{SEP}, it is possible to reexpress (\ref{pe}) in terms of the MGF of the  \text{SINR} as
\begin{equation}
{\cal B}=\frac{\beta_M}{\pi} \sum_{p=1}^{\tau_M}\int_{0}^{\pi/2}M_{\text{SINR}}\left(\frac{a_p^{2}}{2 \sin^{2}(\theta)}\right) d\theta.
\end{equation}
Under shadowed $\kappa$-$\mu$ fading,  substituting the SINR MGF by its expression in (\ref{MGFSKMU}) and swapping the integration order gives
\begin{eqnarray}
{\cal B}^{S\kappa\mu}(\lambda,\alpha)&=&\frac{\beta_M}{2}- \frac{A \beta_M\Gamma(\mu+1)}{2\pi B^{\mu+1}}\sum_{p=1}^{\tau_M}a_p^{2}\int_{0}^{\infty}\!\!\!\left(\!\!\int_{0}^{\pi/2} \frac{{\rm \Psi_1}\left(\mu+1, m; 2,\mu;\frac{ -a_p^{2} \xi \Omega}{2 \sin^{2}(\theta) \mu(1+\kappa)}, \frac{\mu \kappa}{\mu \kappa+m}\right)}{\sin^{2}(\theta)}d\theta\!\!\right)\!\!\nonumber \\&&{\cal E}_{r} \!\!\left[\exp(-\xi r^{\alpha}\sigma^{2} ){\cal L}_{I}^{S\kappa\mu}\left(\xi r^{\alpha} \right)\right] d\xi,
\label{B1}
\end{eqnarray}
Denote by  $\Upsilon$ the inner integral in the RHS of (\ref{B1}) and let $t=\sin^{2}(\theta)$.  Then after some manipulation one obtains
\begin{equation}
\Upsilon=\frac{1}{2}\int_{0}^{1} \frac{{\rm \Psi_1}\left(\mu+1, m; 2,\mu;\frac{- a_p^{2} \xi \Omega}{ 2 t\mu(1+\kappa)}, \frac{\mu \kappa}{\mu \kappa+m}\right)}{t^{\frac{3}{2}} \sqrt{1-t}}dt.
\label{I1}
\end{equation}
To solve $\Upsilon$  we recall the  single integral representation of the Humbert function ${\rm \Psi_1}(a; b; c, c'; z,w )$, for $|w|<1$,   given in  \cite{humbert} as
\begin{equation}
{\rm \Psi_1}\left(a,b; c,c';w , z\right)=\frac{\Gamma(c)}{\Gamma(b)\Gamma(c-b)}\int_{0}^{1}t^{b-1}(1-t)^{c-b-1}(1-t w)^{-a}{\rm {}_{1}F_{1}}\left(a;c',\frac{z}{1-tw}\right) dt,
\label{psi1}
\end{equation}
Substituting ${\rm \Psi_1}$ by its integral representation in (\ref{I1}) and resorting to
\begin{equation}
\int_{0}^{1}\frac{{\rm {}_{1}F_{1}}\left(a+1, b+1;- \frac{c}{x}\right)}{x^{3/2} \sqrt{1-x}}dx =\sqrt{\frac{\pi}{c}}\frac{\Gamma(a+\frac{1}{2})\Gamma(b+1)}{\Gamma(1+a)\Gamma(b+\frac{1}{2})}{\rm {}_{1}{F}_{1}}\left(a+\frac{1}{2}, b+\frac{1}{2};-c\right),
\end{equation}
we obtain
\begin{equation}
\Upsilon=\frac{2\sqrt{2}  \Gamma(\mu+\frac{1}{2})\sqrt{\frac{\mu(1+\kappa)}{\Omega}}}{\Gamma(\mu+1)\sqrt{\xi}}{\rm \Psi_1}\left(\mu+\frac{1}{2}, m; \frac{3}{2},\mu;\frac{ -a_p^{2} \xi\Omega}{2\mu(1+\kappa)}, \frac{\mu\kappa}{\mu\kappa+m}\right).
\label{I2}
\end{equation}
Substituting $\Upsilon$ by its expression in (\ref{B1}) yields the desired result after some simplifications.

\subsection{Proof of ${\cal B}^{S\kappa\mu, \infty}(\alpha)$}
Carrying out the change of variable $x=\Omega \xi $ in (\ref{hkappamu}) and using   \cite[ Ch. 5, Eq. 2]{eldery},  we have
\begin{equation}
{\rm {}_{2}F_{1}}\left(-\frac{2}{\alpha}, m_I ; 1-\frac{2}{\alpha}; -\frac{\Omega_I}{\Omega m_I} x\right) \underset{\frac{\Omega_I}{\Omega} \rightarrow 0}{\approx} 1+\frac{2 \Omega _I}{\Omega(\alpha-2)}x,
\label{appI}
\end{equation}
Subsequently, the following integral arises from  (\ref{hkappamu}):
\begin{equation}
{\cal J}=\int_{0}^{\infty} \frac{{\rm \Psi_1}\left(\mu+\frac{1}{2}, m; \frac{3}{2},\mu;\frac{ -a_p^{2} x}{2\mu(1+\kappa)}, \frac{\mu\kappa}{\mu\kappa+m}\right)}{\sqrt{x}(1+\frac{2 \Omega _I}{\Omega(\alpha-2)}x)}dx.
\label{iref}
\end{equation}
To solve ${\cal J}$, we  introduce the  integral representation of the Humbert function ${\rm \Psi_1}$ given by \cite{humbert}
\begin{equation}
{\rm \Psi_1}\left(a,b; c,c';w , z\right)=\frac{\Gamma(c')}{\Gamma(a)}z^{\frac{1-c'}{2}}\int_{0}^{\infty}t^{a-(c'+1)/2}e^{-t}{\rm I}_{c'-1}(2\sqrt{ t z}){\rm {}_{1}F_{1}}\left(b;c, t  w \right) dt, \quad {\rm Re} (a)>0, |w|<1.
\label{psi11}
\end{equation}
Then substituting  (\ref{psi11}) into ${\cal J}$ and swapping the integration order generate an integral of the form
\begin{equation}
\int_{0}^{\infty}\frac{{\rm I}_{\frac{1}{2}}\left(\sqrt{\frac{ - 2 a_p^{2} x t }{\mu(1+\kappa)}}\right)}{x^{3/4}\left(1+\frac{2 \Omega _I}{\Omega(\alpha-2)}x\right)}dx= \sqrt{\pi}\left(-\frac{a_p^{2}  t }{2\mu(1+\kappa)}\right)^{-\frac{1}{4}}\left(1-e^{-\frac{\sqrt{\frac{ a_p^{2}  t }{\mu(1+\kappa)}}}{\sqrt{\frac{ \Omega _I}{\Omega(\alpha-2)}}}}\right).
\end{equation}
Let $\delta=\frac{\Omega(\alpha-2)}{\Omega_I}$, then resorting to the representation $e^{z} =\sqrt{\frac{ z \pi}{2 }}\left({\rm I}_{\frac{1}{2}}(z)+{\rm I}_{-\frac{1}{2}}(z)\right)$  and using \cite[Eqs. (7.621.4), (9.121.1)]{grad}, we obtain ${\cal J}$, after several manipulations, as
\begin{eqnarray}
{\cal J}&=& \sqrt{2\pi}\frac{\Gamma(\mu)\Gamma(\frac{3}{2})\sqrt{\mu(1+\kappa)}}{\Gamma(\mu+\frac{1}{2})a_p}\left(\frac{m}{\mu \kappa+m}\right)^{-m}+
 \frac{\pi \delta^{\frac{1}{2}}}{\sqrt{2}}{\rm \Psi_1}\left(\mu+\frac{1}{2}, m; \frac{3}{2},\mu;\frac{a_p^{2}\delta }{4\mu(1+\kappa)}, \frac{\mu\kappa}{\mu\kappa+m}\right)\nonumber \\ && -\frac{\pi\Gamma(\mu)\sqrt{\mu(1+\kappa)}}{\sqrt{2} a_p\Gamma(\mu+\frac{1}{2})}{\rm \Psi_1}\left(\mu, m; \frac{1}{2},\mu;\frac{a_p^{2}\delta }{4\mu(1+\kappa)}, \frac{\mu\kappa}{\mu\kappa+m}\right).
\label{iref}
\end{eqnarray}
Tacking all these facts into consideration yields the desired result after some manipulations.
\subsection{Proof of ${\cal B}^{m, \infty}(\alpha)$}
From (\ref{pem}) and (\ref{appI}), the high SIR BEP under Nakagami-$m$ fading involves an integral of the form
\begin{equation}
{\cal K}=\int_{0}^{\infty}\frac{{\rm {}_{1}F_{1}}\left(m+\frac{1}{2}, \frac{3}{2};\frac{ -a_p^{2} \xi}{2m}\right)}{\sqrt{\xi}\left(1+\frac{2 \Omega _I}{\Omega(\alpha-2)}\xi\right)} d\xi,
\end{equation}
which can be solved  using \cite[Eqs. 7.623.1, 9.210.2]{grad}  after recognizing that ${\rm {}_{1}F_{1}}\left(m+\frac{1}{2}, \frac{3}{2};\frac{ -a_p^{2} \xi}{2m}\right)\!=\!e^{-\frac{ a_p^{2} \xi}{2m}}{\rm {}_{1}F_{1}}\left(1-m, \frac{3}{2};\frac{ a_p^{2} \xi}{2m}\right)$. We then obtain
\begin{equation}
{\cal K}=-\frac{\pi \Gamma(m)\sqrt{m}}{\sqrt{2} a_p \Gamma(m+\frac{1}{2})}\left({\rm {}_{1}F_{1}}\left(m, \frac{1}{2};\frac{ a_p^{2}\Omega(\alpha-2) }{4 m \Omega _I }\right)-1\right)+\frac{\pi {\rm {}_{1}F_{1}}\left(m+\frac{1}{2}, \frac{3}{2};\frac{ a_p^{2}\Omega(\alpha-2) }{4 m \Omega _I }\right)}{\sqrt{\frac{2 \Omega _I}{\Omega(\alpha-2)}}}.
\end{equation}
Substituting the latter result in ${\cal B}^{m,\infty}(\alpha)$ and resorting to  \cite[ Eq. (9.210.2)]{grad} completes the proof.

\end{document}